\documentclass[entropy,accept,article,pdftex,moreauthors]{mdpi}

\usepackage{booktabs}
\firstpage{1} 
\pubvolume{1}
\issuenum{1}
\articlenumber{0}
\pubyear{2023}
\copyrightyear{2023}
\datereceived{ } 
\daterevised{ } 
\dateaccepted{ } 
\datepublished{ } 
\hreflink{https://doi.org/} 

\newcommand{\bd}{\boldsymbol}

\newcommand{\dedication}[1]{
\vspace{6pt}\noindent{\fontsize{9}{11.2}\selectfont\textbf{Dedication:} {#1}\par}}

\Title{Learning Traveling Solitary Waves Using Separable Gaussian Neural Networks}

\TitleCitation{Title}

\Author{Siyuan Xing$^{1}$\orcidA{}, Efstathios G. Charalampidis $^{2}$\orcidA{}}

\AuthorNames{Siyuan Xing, Stathis Charalampidis}

\AuthorCitation{Xing, S. Y.; Charalampidis, S.}

\address{%
$^{1}$ \quad Department of Mechanical Engineering, California Polytechnic State University, San Luis Obispo, CA, 93407-0403; sixing@calpoly.edu\\
$^{2}$ \quad Mathematics Department, California Polytechnic State University, San Luis Obispo, CA, 93407-0403; echarala@calpoly.edu}

\corres{Correspondence: sixng@calpoly.edu, echarala@calpoly.edu}

\abstract{In this paper, we apply a machine-learning approach to learn traveling solitary
waves across various families of partial differential equations (PDEs).~Our approach integrates
a novel interpretable neural network (NN) architecture, called Separable Gaussian Neural Networks (SGNN)
into the framework of Physics-Informed Neural Networks (PINNs).~Unlike the traditional PINNs that
treat spatial and temporal data as independent inputs, the present method leverages wave characteristics to
transform data into the so-called co-traveling wave frame.~This adaptation effectively addresses
the issue of propagation failure in PINNs when applied to large computational domains.~Here,
the SGNN architecture demonstrates robust approximation capabilities for single-peakon, multi-peakon,
and stationary solutions within the (1+1)-dimensional, $b$-family of PDEs.~In addition, we expand
our investigations, and explore not only peakon solutions in the $ab$-family but also compacton
solutions in (2+1)-dimensional, Rosenau-Hyman family of PDEs.~A comparative analysis with MLP reveals
that SGNN achieves comparable accuracy with fewer than a tenth of the neurons, underscoring its efficiency
and potential for broader application in solving complex nonlinear PDEs.}


\begin{document}


\section{Introduction}
Physics-informed Neural Networks (PINNs)~\cite{RaissiPINN2019, KarniadakisNature2021} have emerged
as a promising data-driven approach to solving partial differential equations (PDEs) by synthesizing
data and physical laws.~Moreover, they have received considerable traction because they can be efficiently
adapted to solving PDEs defined on domains with arbitrary geometry.~Remarkable results with PINNs have
been achieved across multiple domains and physical situations, such as heat transfer~\cite{CaiJHT2021},
Navier-Stokes~\cite{JinJCP2021} and Euler equations~\cite{PhysRevLett.130.244002}, nonlinear
dynamical lattices~\cite{WeiZhuSPINN_2022,SAQLAIN2023107498}, and medical image processing~\cite{vanHerten2022},
to name a few.

\begin{figure}
\centering
\includegraphics[width=0.95\textwidth]{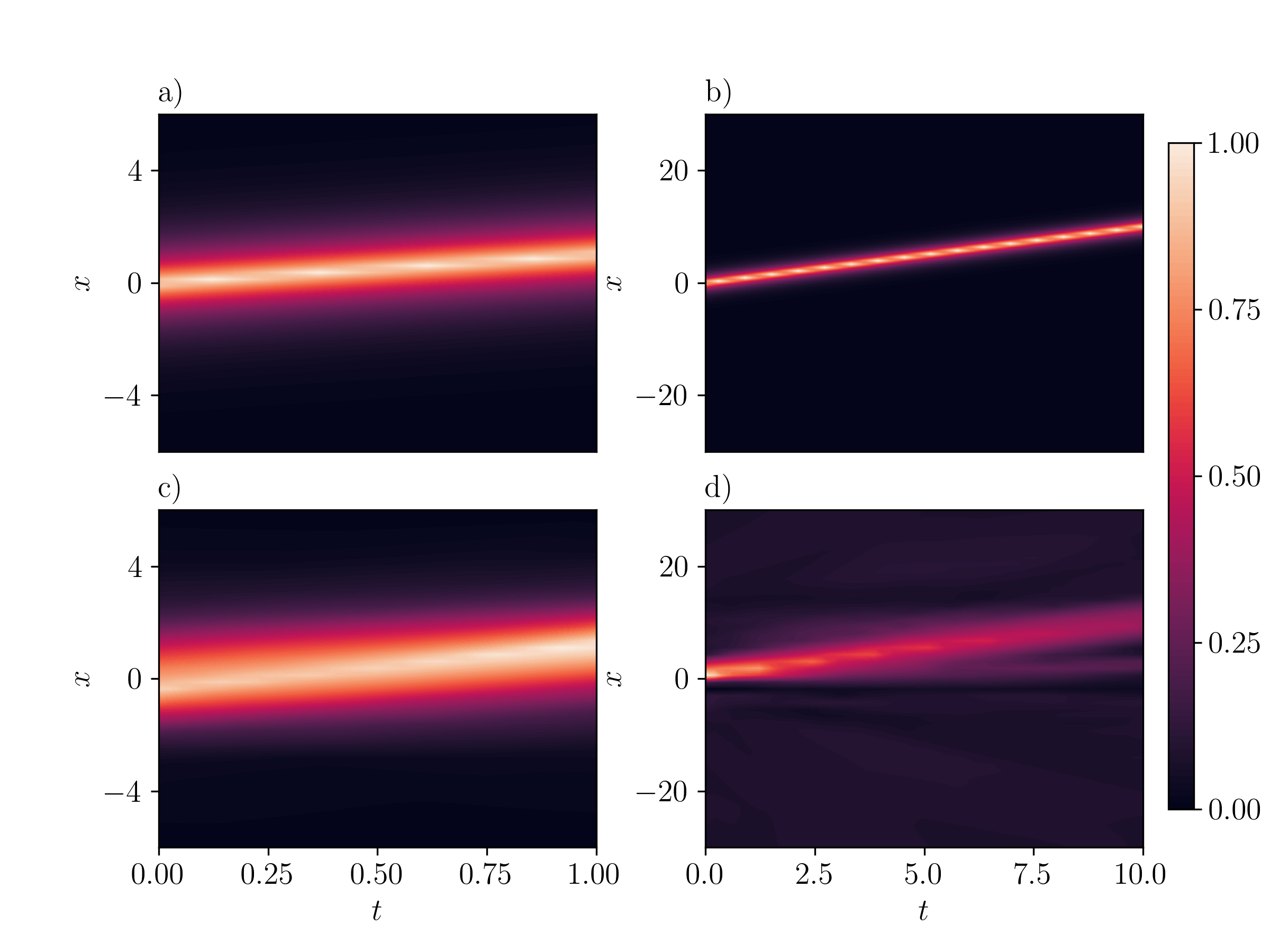}
\caption{Computational results of the spatio-temporal evolution of a peakon in the
Camassa-Holm (CH) equation using PINNs.~The propagation failure of PINNs occurs when
a large spatio-temporal domain, e.g., $[-30,30]\times[0,10]$ is utilized.~In smaller
domains, e.g., $[-5,5]\times[0,1]$ (left panel), PINNs are able to roughly capture
the correct solution.~However, in larger domains (right panel), it tends to converge
to trivial solutions. a) and b): ground truth; c) and d): NN approximation.~The configuration
of the loss function is referred from \cite{WangandYanPhysicaD2021}.
}
\label{fig:pathologies}
\end{figure}

However, many examples of PINNs are limited to "toy" problems situated in low-dimensional spaces with
small spatio-temporal, i.e., computational domains.~It has been observed that PINNs often converge to
incorrect or trivial solutions across a broad spectrum of problems%
~\cite{Wangeltl2021, WangJCP2022, krishnapriyan2021characterizing} (see also%
~\cite{WeiZhuSPINN_2022} for a case where they fail to respect symmetries).~This issue becomes more pronounced
in problems with larger domains, where a phenomenon known as propagation failure~\cite{daw2023mitigating}
frequently occurs.~This challenge arises because PINNs utilize an unsupervised learning scheme to solve PDEs
by minimizing the residual errors of the underlying governing equations.~The presence of this phenomenon does
not ensure convergence to an accurate solution, as numerous trivial solutions can also exhibit zero residuals.%
~Therefore, as the learning process attempts to extend the solution from the initial and/or boundary conditions %
to the interior points, it often becomes "trapped" in trivial solutions.~This phenomenon is particularly common
when PINNs are applied to solve problems with large domains.~Indicatively, Fig.~\ref{fig:pathologies} highlights
this issue in the Camassa-Holm (CH) equation~\cite{PhysRevLett.71.1661} for a single-peakon solution.

To address the pathologies of PINNs, multiple methods have been developed including ones that consider embedding
Fourier features~\cite{tancik2020fourfeat}, adaptive sampling~\cite{WuCMAME2023, daw2023mitigating},
and those respecting causality~\cite{wang2024respectingcausality}.~In fact, besides the physical laws embodied
in PDEs themselves, the mathematical properties of their solutions can be leveraged too.~For example, traveling
waves (TWs) to PDEs are solutions of the form $u(x\pm ct)$, where $c$ is their speed (the ``-'' and ``+'' signs
correspond to TWs moving, i.e., traveling to the right and left, respectively).~However, few efforts have been
devoted to embedding such mathematical properties of solutions into PINNs (see,~\cite{WeiZhuSPINN_2022} for
the development of symmetry-preserving PINNs) such that the output of neural networks (NNs) will automatically
respect the corresponding features of the solution.~This is expected to improve the efficiency in training
and increase the opportunity for NNs to converge to correct solutions.

In this paper, we aim to enhance PINNs by pursuing this route.~More specifically, we will focus
on seeking TW solutions to nonlinear PDEs using PINNs with input transformed into a frame that
co-moves with the solution, i.e., co-traveling frame.~This idea has been explored in the recent
work in~\cite{CINN_Barga_Neto_2023} in which the characteristics of hyperbolic PDEs are encoded
in the network by adding a characteristic layer.~Herein, we will use this structure to learn
TWs, i.e., solitary waves in multiple families of one and two-dimensional nonlinear PDEs.~Those include
the $b$- and $ab$-families of peakon equations~\cite{HOLM2003437,Alexandrou-Himonas-ab-family-2016}
which contain the (completely integrable) Camassa-Holm (CH) and Degasperis-Procesi equations~\cite{PhysRevLett.71.1661,Degasperis2002}
(see, also~\cite{FUCHSSTEINER198147}), and the Rosenau-Hyman compacton equations~\cite{Rosenau_compacton_2007}
(see, also~\cite{Rosenau2000}).~In addition, a novel interpretable NN -- Separable
Gaussian Neural Networks (SGNNs)~\cite{XingandSunSGNN2023} -- will be employed to extract solution
forms in the sense of generalized Gaussian radial-basis functions.~The description of this network
will be deferred to Section 2, along with the discussion about its advantages.

The rest of the paper is organized as follows.~In Section 2, we introduce the architecture of SGNN
with its input transformed to co-traveling frames.~Our aim is to integrate this structure into the
framework of PINNs, and identify TWs to the nonlinear PDEs of interest in this work.~In Sections 3 and
4, we demonstrate the applicability of the method to the study of peakons in the $b$- and $ab$-families
of peakon equations, respectively.~Then, we extend this approach in Section 5 to identify 2D compacton
configurations.~We mention in passing that the architecture can easily predict such higher-dimensional
solutions which have not been studied in the realm of PINNs to the best of our knowledge.~At last, we
perform an extensive comparison of the two different architectures of PINNs with different network
structures in Section~6 where the advantages and disadvantages, as well as limitations of SGNN are discussed%
.~We conclude our findings in Section 7, and present future research directions.

\section{Methods}

\subsection{Architecture of SGNN for traveling waves}

\begin{figure}
    \centering
    \includegraphics[width=0.95\textwidth]{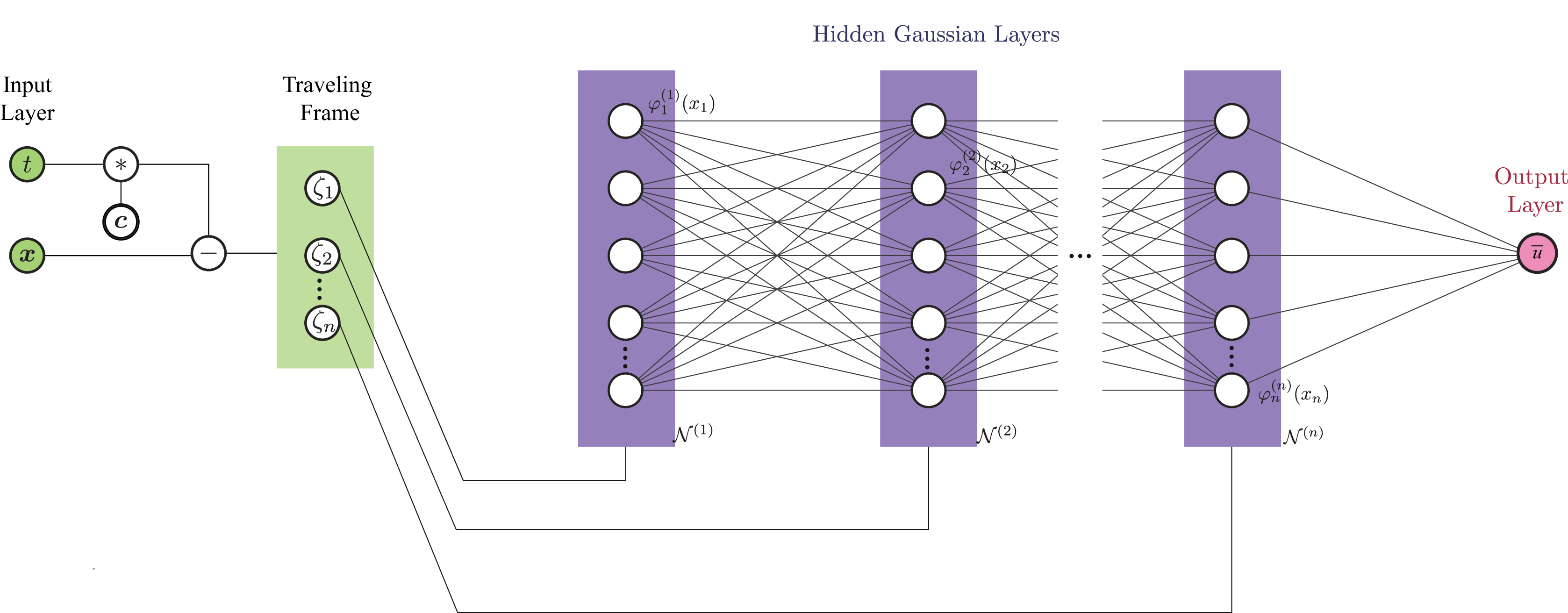}
    \caption{The architecture of SGNN with input transformed into the co-traveling frame
    whose coordinates are $\bd{\zeta}=\bd{x}-\bd{c} t$, where $\bd{c}$ represents the
    velocity of the wave.~Different from MLPs, the transformed input is then
    split and fed sequentially to hidden layers of an SGNN that consist of univariate
    functions.~The multiplication and addition of such univariate functions in feedforward
    propagation can eventually lead to the summation of a set of multivariate functions
    used to approximate the solution of a PDE.}
    \label{fig:SGNN}
\end{figure}

Inspired by~\cite{CINN_Barga_Neto_2023}, a $d$-dimensional (in space) TW is mapped into
a frame that co-moves with it by performing the following coordinate transformation
\begin{equation}
    \zeta_i=x_i-c_i t,~~i=1,2,\dots, d,
\end{equation}
where $c_i$ is its constant velocity in the $i$-th dimension.~Under such a transformation,
a TW becomes a stationary wave in the co-traveling frame.~As shown in Fig. \ref{fig:SGNN},
the coordinates $\zeta_i$ ($i=1,2,\dots,d)$ become the input of the SGNN.~The coordinates
are then divided according to their dimensions, and sequentially fed to the feedforward layers.%
~This results in a number of layers that is equal to the number of spatial dimensions.%
~The neurons of each layer we consider are generalized univariate Gaussian functions in
the form
\begin{equation}
    {\varphi}(\zeta_i,{\mu},{\sigma})=exp\left((\zeta_i-\mu)^\alpha/\sigma^2\right),
\end{equation}
where $\alpha\in \mathcal{R}-\{0\}$. When $\alpha=2$, $\varphi$ is the regular univariate
radial-basis Gaussian function.~In this paper, we will adopt $\alpha=1$ for peakon solutions,
and $\alpha=2$ for other solutions.

The first hidden layer of SGNN receives a single input: the partial coordinate $\zeta_1$.%
~Subsequent hidden layers take two inputs - the output from the preceding hidden layer and a
coordinate in the traveling frame.~The network culminates in a dense output layer, which
aggregates the outputs from the final hidden layer.~The mathematical representation of
SGNN~\cite{XingandSunSGNN2023} is defined in the form
\begin{align}
\mathcal{N}^{(1)}_i&={\varphi}^{(1)}_i(\zeta_1,{\mu}^{(1)}_i,{\sigma}^{(1)}_i), ~ 1\leq i \leq N_1, \label{Eq: sec_4: layer 1}\\
\mathcal{N}^{(\ell)}_i&={\varphi}^{(\ell)}_i(\zeta_\ell,{\mu}^{(\ell)}_i,{\sigma}^{(\ell)}_i)\sum_{j=1}^{N_l} {W}^{(\ell)}_{ij}\mathcal{N}^{(\ell-1)}_j, ~~2\leq \ell\leq d, \,\,\, 1\leq i \leq N_l, \label{Eq:sec_4: layer l}\\
\bar{u}(\mathbf{x})&=\mathcal{N}(\mathbf{x})=\sum_{j=1} ^{N_d}W_j^{(d)}\mathcal{N}^{(d)}_j, \label{Eq:sec_4: last layer}
\end{align}
where $\mathcal{N}^{(l)}_i$ represents the output of the $i$-th neuron of the $l$-th layer,
$N_l$  stands for the number of neurons of the $l$-th layer, and $\bar{u}$ is the output of
SGNN.~When $d>2$, the weights of the output layer are set to 1.

Thanks to the separable property of Gaussian radial-basis functions, the forward propagation
of such univariate Gaussian functions yields the summation of multiple chains of univariate
Gaussian functions which are equivalent to the summation of high dimensional Gaussian
radial-basis functions.~In other words, the output of an SGNN is equivalent to the output of
a GRBFNN~\cite{park_universal_1991} in the form of
\begin{equation}
\bar{u}(\mathbf{x})=\sum_{j=1}\mathcal{W}_jG_j, \label{Eq:sec_2: SGNN-GRBFNN}
\end{equation}
where $G_j$ is a $d$-dimensional Gaussian radial-basis function.

The SGNN offers several advantages.~Firstly, it is interpretable.~The parameters of a
neuron depict its local geometrical information (center and width).~Without the composition
of nonlinear activation functions, a human-interpretable explicit output form of
Eq.~\ref{Eq:sec_2: SGNN-GRBFNN} can be obtained, in the sense of Gaussian radial-basis
functions.~Secondly, SGNN is easier to tune than MLP. This is because the depth of SGNN
is identical to the number of dimensions; therefore, the only tunable hyperparameter is
the width of each layer. Lastly, it can achieve several-order-of-magnitude more accurate
results than MLP when approximating complex functions.~The interested reader can consult%
~\cite{XingandSunSGNN2023} for detailed comparisons between SGNN and MLP.

\subsection{Physics-informed Machine Learning}
The SGNN is adopted to approximate the solution $u(x,t)$ of PDEs in the form
\begin{equation}
    u_t + \mathcal{F}[u,u_x,u_{xx},\dots]=0,
\end{equation}
which is subject to boundary and initial conditions (abbreviated hereafter as
BCs and ICs, respectively)
\begin{align}
    \mathcal{B}[u]&=0,\\
    u(x,0)&=f(x).
\end{align}
The loss function is defined as
\begin{equation}
    \mathcal{L}=\lambda_r\mathcal{L}_{r} +  \lambda_{bc}\mathcal{L}_{bc} + \lambda_{ic}\mathcal{L}_{ic},
\end{equation}
where
\begin{align}
        \mathcal{L}_{r}&=\frac{1}{N_r}\sum_{i=1}^{N_r}\left|\mathcal{R}(\bd{x}_r^i,t_r^i)\right|^2,\label{eq:PDE residual error}\\
        \mathcal{L}_{bc}&=\frac{1}{N_{bc}}\sum_{i=1}^{N_{bc}}\left|\mathcal{B}[u](\bd{x}_{bc}^{i}, t_{bc}^i)\right|^2,\label{eq: bc error}\\
        \mathcal{L}_{ic}&=\frac{1}{N_{ic}}\sum_{i=1}^{N_{ic}}\left|u(x_{ic}^i,0)-f(x_{ic}^i)\right|^2,\label{eq: ic error}\\
\end{align}
and $\lambda_r$, $\lambda_{bc}$, $\lambda_{ic}$ are scaling factors.~Here,
$\mathcal{L}_{r}$, $\mathcal{L}_{bc}$, and  $\mathcal{L}_{ic}$  represent the
MSE (mean-squared error) of PDEs,  BCs, and ICs, respectively.~The collocation points
denoted as $\{x_r^i,t_r^i\}$ and $\{x_{bc}^i, t_{bc}^i\}$ are randomly sampled in
the domain and on the boundary, respectively.~In addition, $\{x_{ic}^i\}$ are spatial
points sampled at $t=0$.

\subsection{Training scheme}

In the 1D problems presented next, we employ a two-stage training process for SGNN,
initially using the ADAM optimizer~\cite{DBLP:journals/corr/KingmaB14} followed by
the L-BFGS algorithm~\cite{liu1989limited}.~This approach allows us to leverage the
L-BFGS algorithm's capability to enhance convergence accuracy after the preliminary
optimization with ADAM. In contrast, for 2D problems, we solely rely on the ADAM
optimizer due to the computational demands of running L-BFGS with larger datasets.%
~The training dataset is randomly sampled using the 'Sobol' method, which empirically
can yield better results~\cite{WuCMAME2023}.~The validation dataset is created through
a method of even partitioning across the domain and boundaries, ensuring comprehensive
coverage and testing of the model's predictive capabilities.~Throughout the training phase,
the SGNN's weights are initialized based on a uniform random distribution.~Additionally,
the initial centers of the univariate Gaussian neurons are distributed evenly across the respective
dimensions, with their initial widths defined by the distance between adjacent centers.

\section{Peakons in  $b$-family}
The first model-PDE we consider in this work is the well-known $b$-family
of peakon equations:
\begin{equation}
    u_t - u_{xxt} + (b+1)uu_x = bu_xu_{xx}+uu_{xxx},
    \label{Eq:b-family}
\end{equation}
that was introduced in~\cite{HOLM2003437}.~It has been proposed as
model for the propagation of shallow water waves~\cite{HOLM2003437}
with the parameter $b$ related to the Kodama transformation group of
shallow water water equations~\cite{KODAMA1985193,KODAMA1985245}.%
~Moreover, Eq.~\eqref{Eq:b-family} contains two completely integrable
models for $b=2$ and $b=3$, known as the Camassa-Holm equation~\cite{PhysRevLett.71.1661,CAMASSA19941}
(see, also~\cite{FUCHSSTEINER198147}) and Degasperis-Procesi equation~\cite{Degasperis2002},
respectively.%

The striking feature of the $b$-family of Eq.~\eqref{Eq:b-family}
is that it possesses explicit single-peakon
\begin{equation}
    u(x,t)=ce^{-|x-ct|},
    \label{Eq:b-family-peakon-solution}
\end{equation}
and multi-peakon solutions
\begin{equation}
u(x,t)=\sum_{j=1}^{N}p_j(t)e^{-|x-q_j(t)|},
\end{equation}
for all values of $b$, where $q_j$ and $p_j$ are the position and amplitude
of $j$-th peakon with $N$ representing the number of peakons, i.e., $j=1,\dots,N$.%
~The peakon solution of Eq.~\eqref{Eq:b-family-peakon-solution} (and similarly, its
multi-peakon version) is not differentiable at its center, rendering its analytical
and numerical study (from the PDE point of view) an extremely challenging task
(see, \cite{Charalampidis_b_family_peakon_2023} for the spectral stability analysis
of peakons).~It should be noted in passing that alongside the existence of peakon
solutions, the $b$-family possesses explicit stationary solutions known as
"leftons"~\cite{Charalampidis_b_family_peakon_2023} (and references therein)
given by
\begin{equation}
    \label{eq:leftons}
    u=A(\cosh(\gamma(x-x_0)))^{-\frac{1}{r}},~~~ \gamma=-\frac{b+1}{2},
\end{equation}
where $A$ and $x_0$ are their amplitude and center, respectively.~These solutions
also emerge numerically upon propagating Gaussian initial data to Eq.~\eqref{Eq:b-family}
for $b<-1$.~Even more, the propagation of Gaussian initial data to the $b$-family
with $-1<b<1$ results in the emergence of "ramp-cliffs" (see~\cite{Charalampidis_b_family_peakon_2023}
and references therein) That is, solutions that involve a combination of
ramp-like solutions of the Burgers equation (evolving as $x/t$) together
with an exponentially-decaying tail, i.e., cliff.

Having introduced the model of interest, we will use the SGNN to approximate both one-
peakon and multi-peakon solutions in the next section.

\subsection{Single peakon}
\subsubsection{Camassa-Holm $(b=2)$}
We first inspect a one-peakon/one-antipeakon solution in the Camassa-Holm (CH) equation.%
~The computational domain we consider is $\Omega=\lbrace{(x,t):[-20,20]\times[0,10]\rbrace}$.%
~We adopt periodic BCs and ICs $u(x,0)=e^{-|x|}$ (i.e., $c=1$).~For our analysis, we employ
a one-layer SGNN with 60 neurons.~As both centers and widths are trainable, the total number
of trainable parameters is $180$.

\begin{figure}[!ht]
    \centering
    \includegraphics[width=0.9\textwidth]{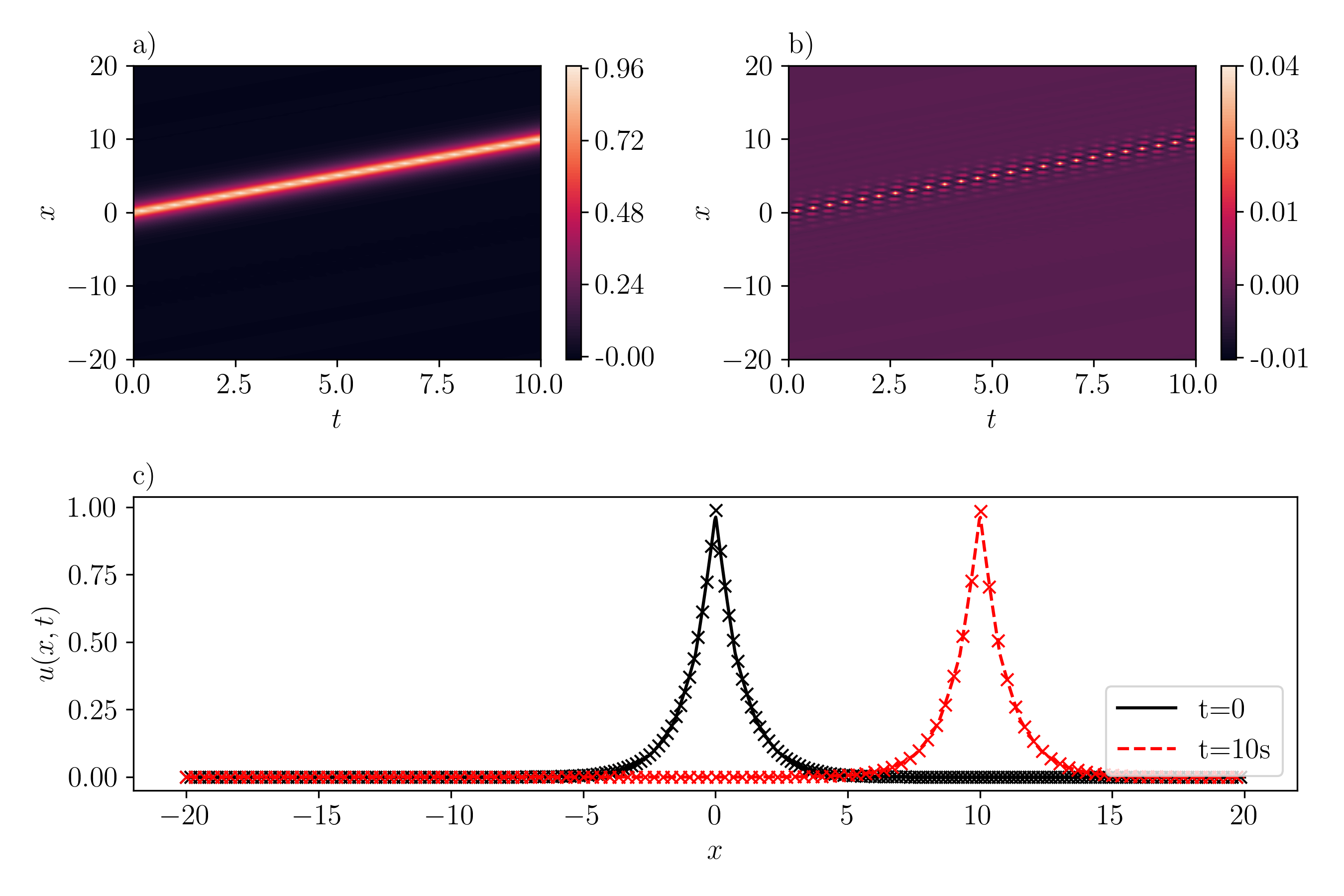}
    \caption{A one-peakon solution in the CH equation ($b=2$) with $c=1$.
    a): $\bar{u}(x,t)$ inferred by SGNN; b): error $e(x,t)=u(x,t)-\bar{u}(x,t)$;
    c): $\bar{u}(x,t)$ at two time instants.~In c), "x" markers represent the exact
    solution while lines represent the prediction by SGNN.~The training loss is
    $8.43e-3$, with $\lambda_{ic}=1,000$, $\lambda_{bc}=1$.~Validation error:
    $\Vert e \Vert=3.90e-2$, $ \Vert e \Vert_2=7.21e-6$.}

    \label{fig:fig2}
\end{figure}

The data collection process involves the sampling of $2^{12}=4,096$ points within the specified domain.%
~Additionally, we use the 'Sobol' sampling scheme to gather $2^{9}=512$ boundary points, and another
$512$ spatial points satisfying the initial condition.~It should be noted that the number of
samples is larger than the number reported in the literature.~This increase in sample size is
attributed to the comparatively larger domain size in our analysis.~We train SGNN for $5,000$
epochs using ADAM~\cite{DBLP:journals/corr/KingmaB14}, followed by L-BFGS~\cite{liu1989limited}
to refine the results.~The dataset is divided into 8 mini-batches.~The learning rate of ADAM is
$1e-2$ for the first $100$ epochs, and $1e-3$ for the rest.~We report that the mean-squared loss
is $8.43e-3$ when training finishes.~It should also be noted that this value is scaled by a relatively
large scaling factor ($\lambda_{ic}=1000$) that is selected using trial and error.~On the other hand,
the mean-squared validation error is much smaller, with a value of $7.21e-6$.~The maximum absolute
validation error is $3.90e-2$.~As illustrated in Fig.\ref{fig:fig2} (b), the inferred peakon solution
with $c=1.0$ accurately approximates the exact solution with the error getting maximized at the crest
of the peakon.~The good agreement is also demonstrated in Fig.\ref{fig:fig2} (c), where the "x" markers
stand for the exact solution [cf.~Eq.~\eqref{Eq:b-family-peakon-solution}], and line for the predicted
solution by SGNN at two different instant of times (see, the legend in the figure).

\begin{figure}[!ht]
    \centering
    \includegraphics[width=0.9\textwidth]{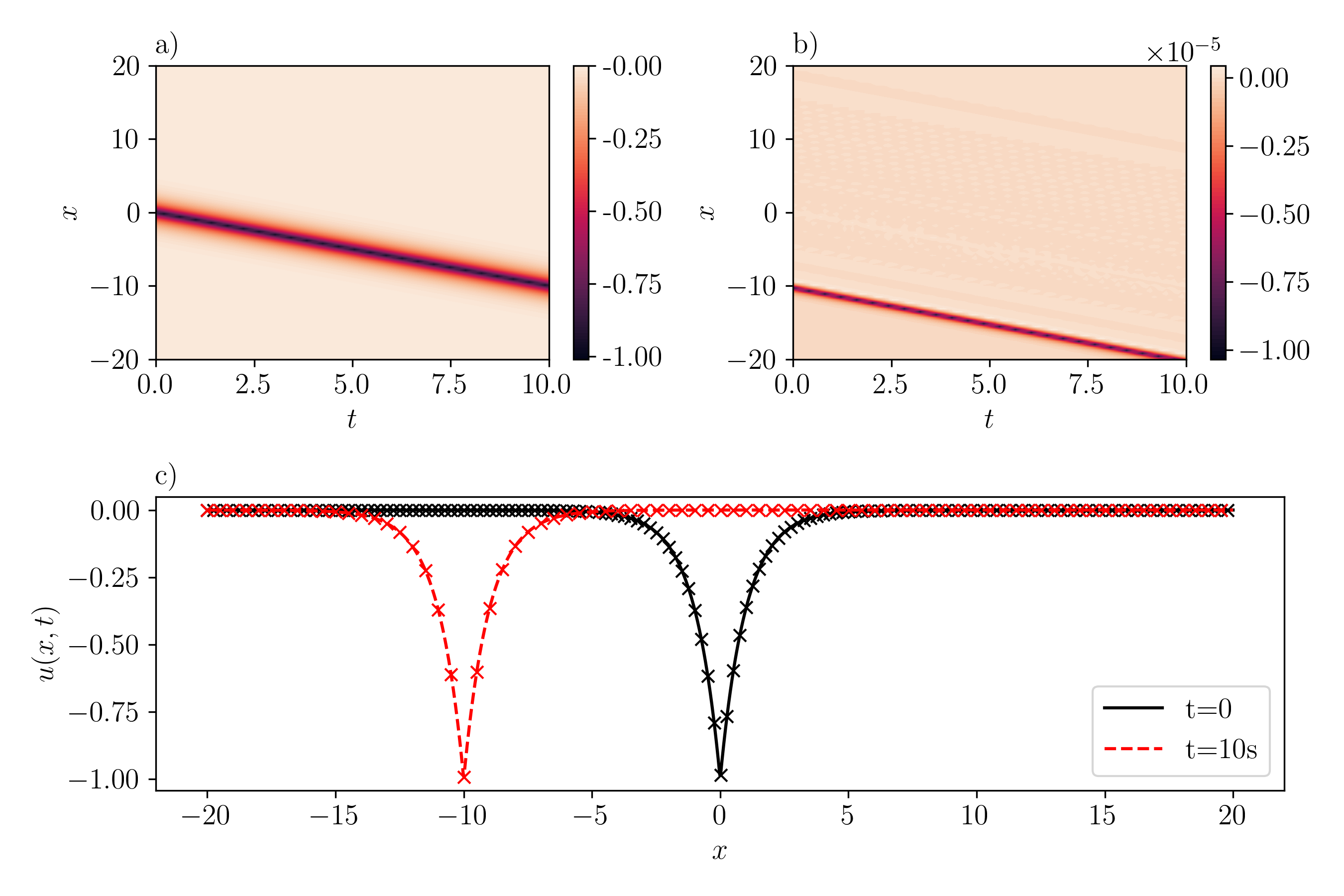}
    \caption{Same as Fig.~\ref{fig:fig2} but for an one-antipeakon solution in the CH equation
    ($b=2$) with $c=-1$.~a): $\bar{u}(x,t)$ inferred by SGNN; b): error $e(x,t)=u(x,t)-\bar{u}(x,t)$;
    c): $\bar{u}(x,t)$ at two time instants. In c), "x" represents the exact solution while
    lines represent the prediction by SGNN.~The training loss is $1.94e-11$, with
    $\lambda_{ic}=1,000, \lambda_{bc}=1$. Validation error:  $\Vert e \Vert=1.02e-5$, $\Vert e \Vert_2=9.59e-13$.}
    \label{fig:fig3}
\end{figure}

A case corresponding to an anti-peakon solution with $c=-1.0$ is represented in Fig. \ref{fig:fig3}.%
~The prediction by SGNN yields a mean-squared loss of $1.94e-11$.~This means that the inference of SGNN
precisely matches the exact solution.~The largest error occurs on the characteristic curve $x+t=-10$,
with the magnitude level of $1e-5$.

\subsubsection{Other values of $b$}
\begin{figure}[!ht]
    \centering
    \includegraphics[width=0.9\textwidth]{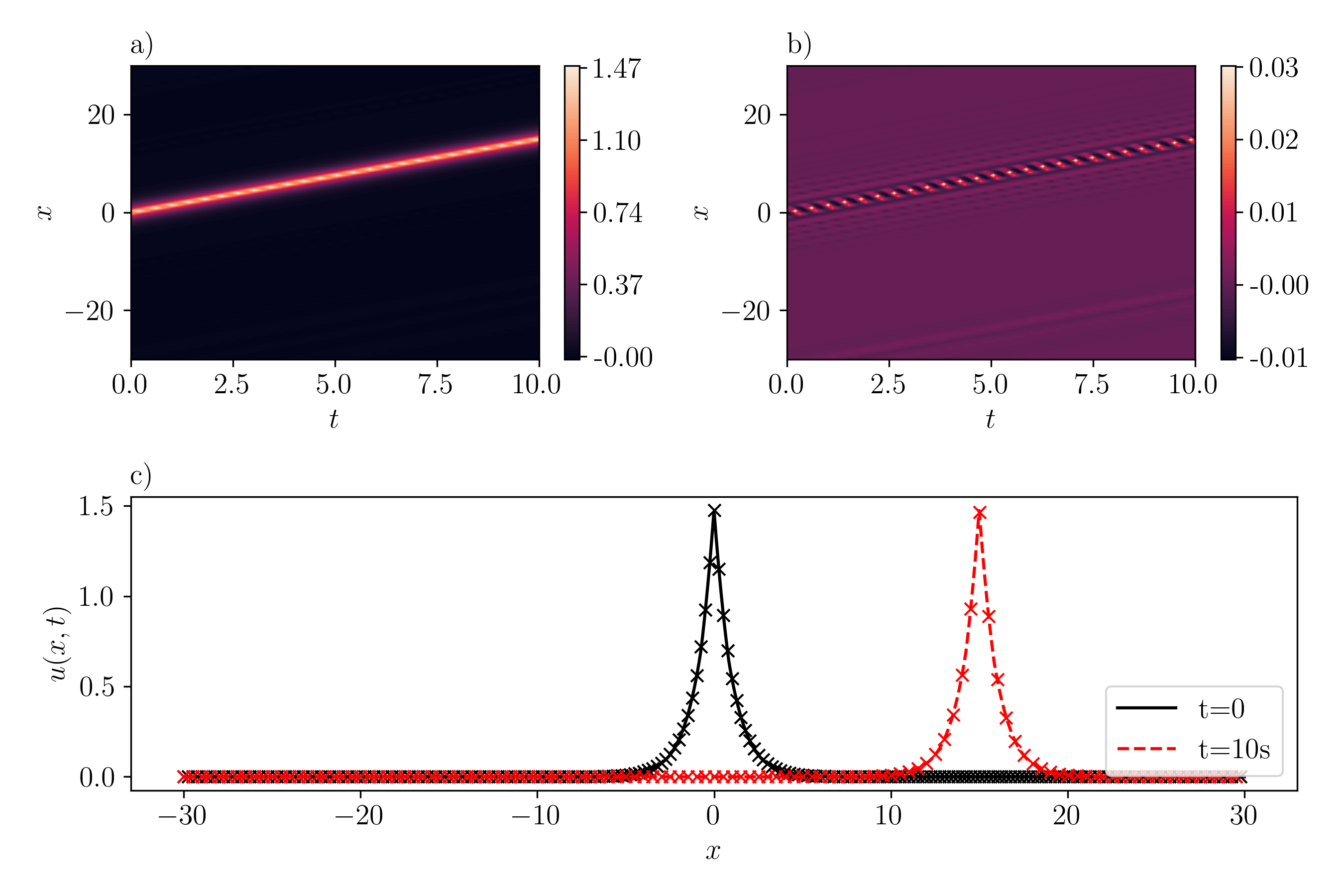}
    \caption{Same as Fig.~\ref{fig:fig3} but for a one-peakon solution of the
    $b$-family with $b=0.8$ and $c=1.5$.~a): $\bar{u}(x,t)$ inferred by SGNN;
    b): error $e(x,t)=u(x,t)-\bar{u}(x,t)$; c): $\bar{u}(x,t)$ at two time
    instants.~In c), "x" represents the analytical solution while curves
    represent the prediction by SGNN. The training loss is $9.18e-2$, with
    $\lambda_{ic}=1,000, \lambda_{bc}=1$. Validation error: $ \Vert e \Vert=2.74e-2$, $ \Vert e \Vert_2=4.09e-6$.}
    \label{fig:fig4}
\end{figure}

\begin{figure}[!ht]
    \centering
    \includegraphics[width=0.9\textwidth]{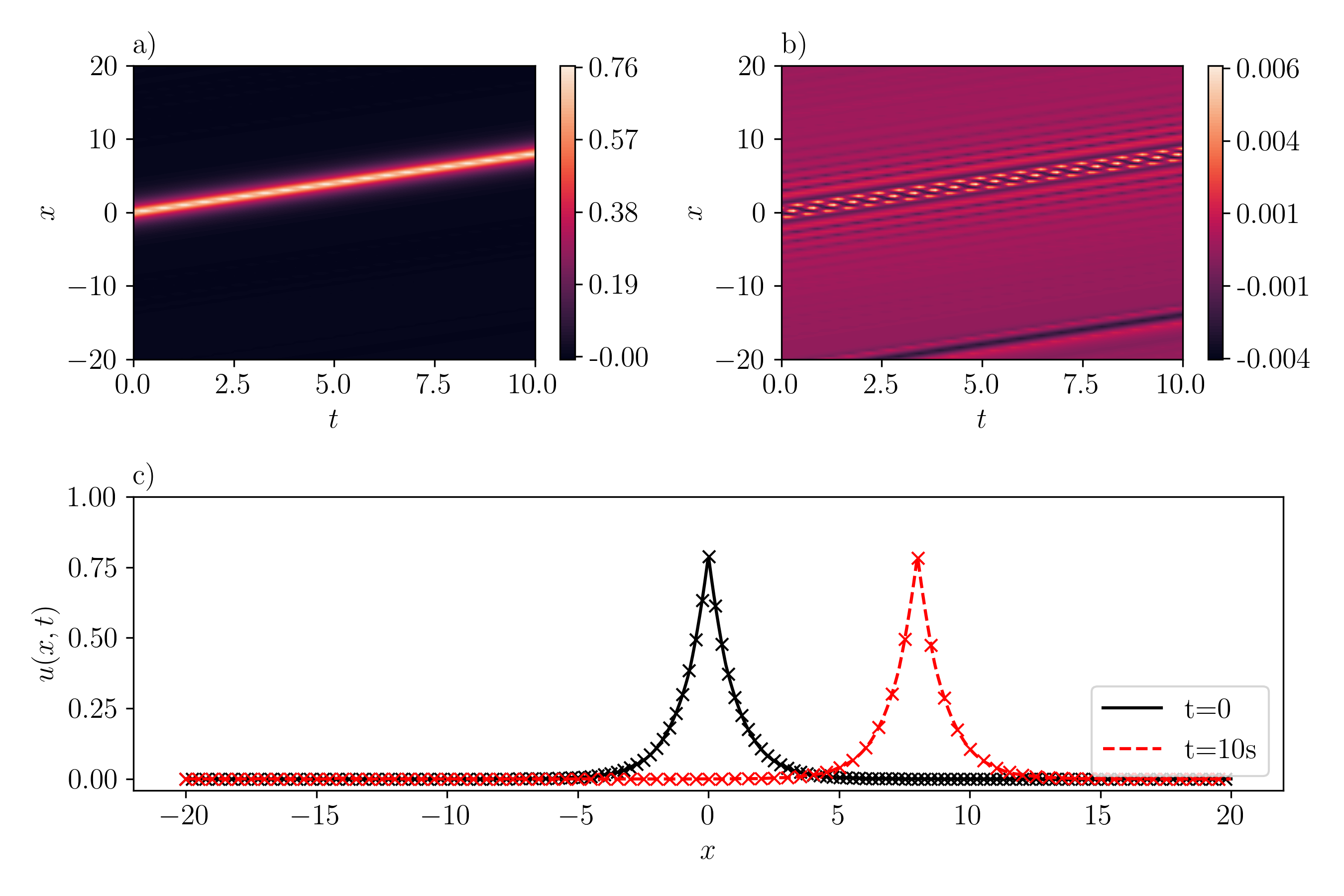}
    \caption{Same as Fig.~\ref{fig:fig4} but with $b=-1.0$ and $c=0.8$.~a):
    $\bar{u}(x,t)$ inferred by SGNN; b): error $e(x,t)=u(x,t)-\bar{u}(x,t)$;
    c): $\bar{u}(x,t)$ at two time instants.~The format in panel c) is the same
    as in panel c) of Fig.~\ref{fig:fig4}.~Here, the training loss is $3.10e-2$,
    with $\lambda_{ic}=10,000$, and $\lambda_{bc}=1$. Validation error:
    $\Vert e \Vert=5.92e-3$, $ \Vert e \Vert_2=3.50e-5$.}
    \label{fig:fig5}
\end{figure}

We next investigate the emergence of peakons using SGNN for different values of $b$.~Indeed, Fig. \ref{fig:fig4} (a)
presents a peakon solution predicted by SGNN with $b=0.8$ and $c=1.5$.~While the temporal domain remains as
$[0,10]$, the spatial domain is enlarged to $[-30,30]$ in order to accommodate the rise of velocity, and thus
the peakon "fits" in the computational domain over its propagation.~As shown in Fig. \ref{fig:fig4} (b), the
prediction matches very well with the exact solution.~The mean-squared validation error is $4.09e-6$, and the
maximum absolute error is $0.0274$.~The maximum absolute error appears at the region where the $u(x,t)$ reaches
its peak value.~The training loss after $5,000$ epochs is reduced to $9.18e-2$.~The waveforms at $t=0$ and $t=10$
are depicted in \ref{fig:fig4} (c), where lines represent SGNN's prediction, and "x" markers represent
the exact solution, respectively.~The predicted peakon solution with $b=-1, c=0.8$ is presented in Fig.~\ref{fig:fig5}.%
~Likewise, a good agreement between inference by SGNN and the exact solution $u(x,t)=0.8e^{-|x-0.8t|}$ is achieved,
with a training loss of $3.1e-2$, a mean-squared validation loss of $3.5e-5$, and a maximum absolute validation
loss of $5.92e-3$.

\begin{figure}[!ht]
    \centering
    \includegraphics[width=0.8\textwidth]{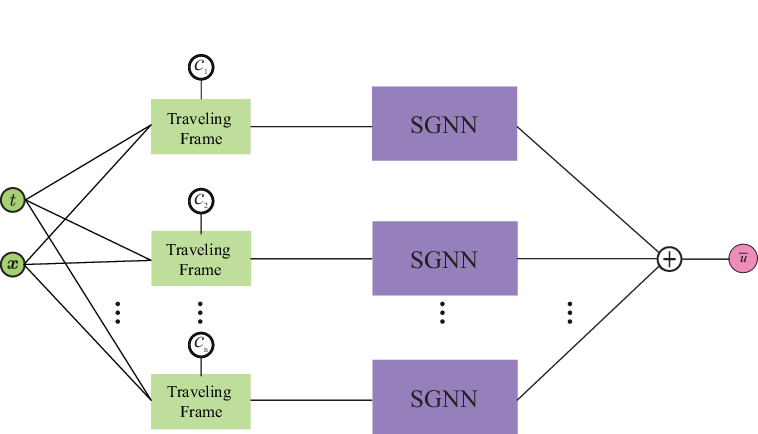}
    \caption{The NN architecture used for the study of multiple peakon configurations.}
    \label{fig:SGNN-multiple-peakon}
\end{figure}

\subsubsection{Interacting peakons}
Having discussed the prediction of single-peakon (and anti-peakon) solutions in the
$b$-family, we now turn our focus to cases involving two-peakon configurations in the
CH equation ($b=2$), thus emulating their interactions.~In particular, we focus on the
following three specific scenarios: (1) peakons traveling along the same direction with
identical speed, (2) peakons traveling in the same direction but at different velocities,
and (3) peakons moving in opposite directions.~Given that peakons can travel at varying
speeds and in distinct directions in space (i.e., either left or right), we employ multiple
SGNNs to approximate these peakons, allocating one SGNN per peakon.~The sum of such SGNNs
produces the NN-solution of Eq.~\ref{Eq:b-family}, and the NN structure in this case is
shown in Fig.~\ref{fig:SGNN-multiple-peakon}.~During the training stage, the loss functions
associated with the PDE and BCs are identical to those in Eqs.~\ref{eq:PDE residual error}
and~\ref{eq: bc error}.~However, it is necessary to modify the loss function of ICs such
that the output of each SGNN at $t=0$ accurately reflects the corresponding peakon solution
at $t=0$.

We inspect the response from $t=0$ to $t=10$, within a spatial domain $[-30, 30]$.~Two
one-layer SGNNs with $40$ neurons are used.~Each training dataset is generated by randomly
sampling $2^{13}=8,192$ collocation points within the domain, and $2^{10}=1024$ points on the
boundary.~The dataset is then divided into $8$ mini batches.~The results are obtained with
$5,000$ training epochs by ADAM, followed by refinement by L-BFGS (as before).~The validation
set is generated by uniformly sampling a 50 $\times$ 100 grid in the domain including BCs and
ICs.

\begin{figure}[!ht]
    \centering
    \includegraphics[width=0.9\textwidth]{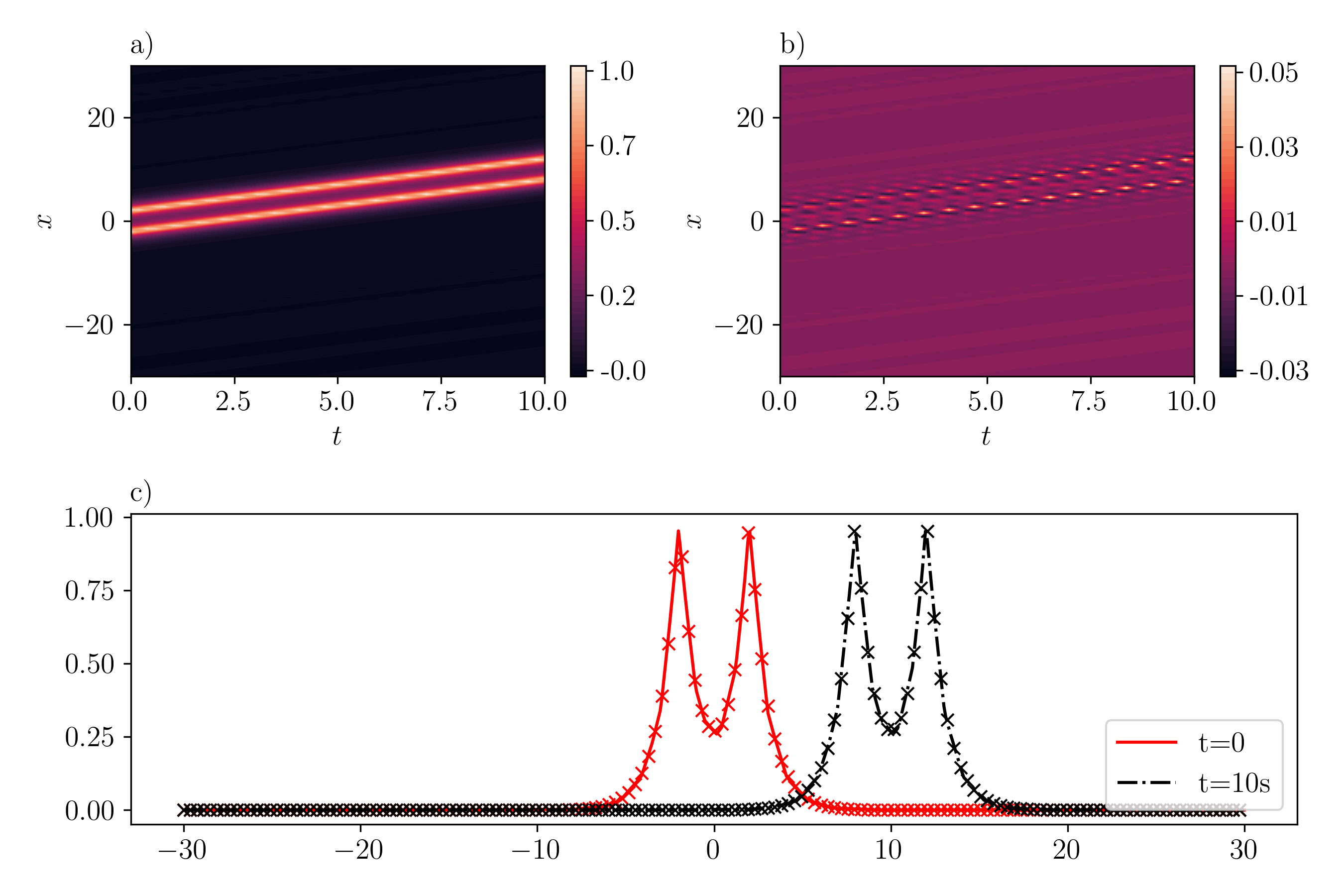}
    \caption{Two peakons with identical traveling speed ($c=1$) in the CH equation.
    a): $\bar{u}(x,t)$ inferred by SGNN; b): error $e(x,t)=u(x,t)-\bar{u}(x,t)$;
    c): $\bar{u}(x,t)$ at two time instants.~In c), "x" represents the exact solution
    while lines represent the prediction by SGNN.~The training loss is $4.27e-3$, with
    $\lambda_{ic}=\lambda_{bc}=100$. Validation error: $\Vert e\Vert=5.22e-2, \Vert e\Vert_2=2.99e-5$.}
    \label{fig:fig6}
\end{figure}

In Fig.~\ref{fig:fig6}, two peakons traveling towards the right with identical speed $c=1$ are
presented.~The ICs employed here are $u(x,0)=e^{-|x+2|}+e^{-|x-2|}$, which forms a bi-nominal
shape.~The training error is $4.27e-3$, with scaling factors $\lambda_{ic}=\lambda_{bc}=100$.%
~As shown in Fig.~\ref{fig:fig6}(a), the peakons maintain their distance during propagation.~Moreover,
it can be discerned from Fig. \ref{fig:fig6}(b) that SGNN is capable of making very good predictions
of such configurations.~Indeed, the mean-squared and maximum absolute errors are $2.99e-5$ and
$5.22e-2$, respectively for this case.~The good agreement between SGNN and exact solutions is
further demonstrated in Fig. \ref{fig:fig6}(c), where "x" makers are for exact solutions and solid
lines are predictions by SGNN.

\begin{figure}[!ht]
    \centering
    \includegraphics[width=0.9\textwidth]{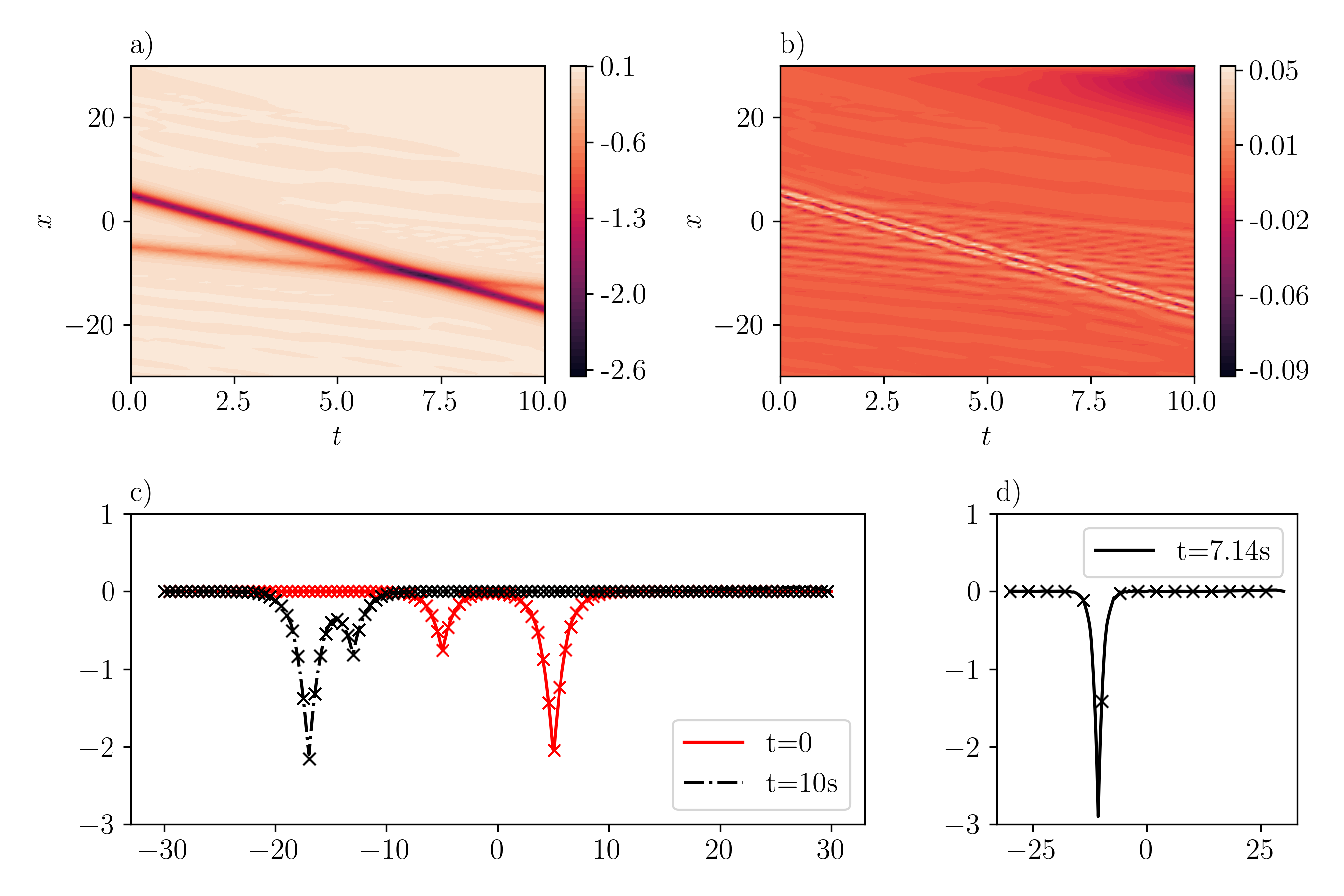}
    \caption{Same as Fig.~\ref{fig:fig6} but for the case corresponding to the
    interaction of two anti-peakons ($c_1=-0.8$, $c_2=-2.2$) in the CH equation.%
    ~a): $\bar{u}(x,t)$ inferred by SGNN; b): error $e(x,t)=u(x,t)-\bar{u}(x,t)$;
    c): $\bar{u}(x,t)$ at two time instants.~The format of panel c) is the same as
    the one in Fig.~\ref{fig:fig6}.~The training loss is $1.88e-2$, with $\lambda_{ic}=\lambda_{bc}=100$.%
    Validation error: $\Vert e\Vert=9.0e-2$, $\Vert e \Vert_2=9.12e-5$.}
    \label{fig:fig7}
\end{figure}

\begin{figure}[!ht]
    \centering
    \includegraphics[width=0.9\textwidth]{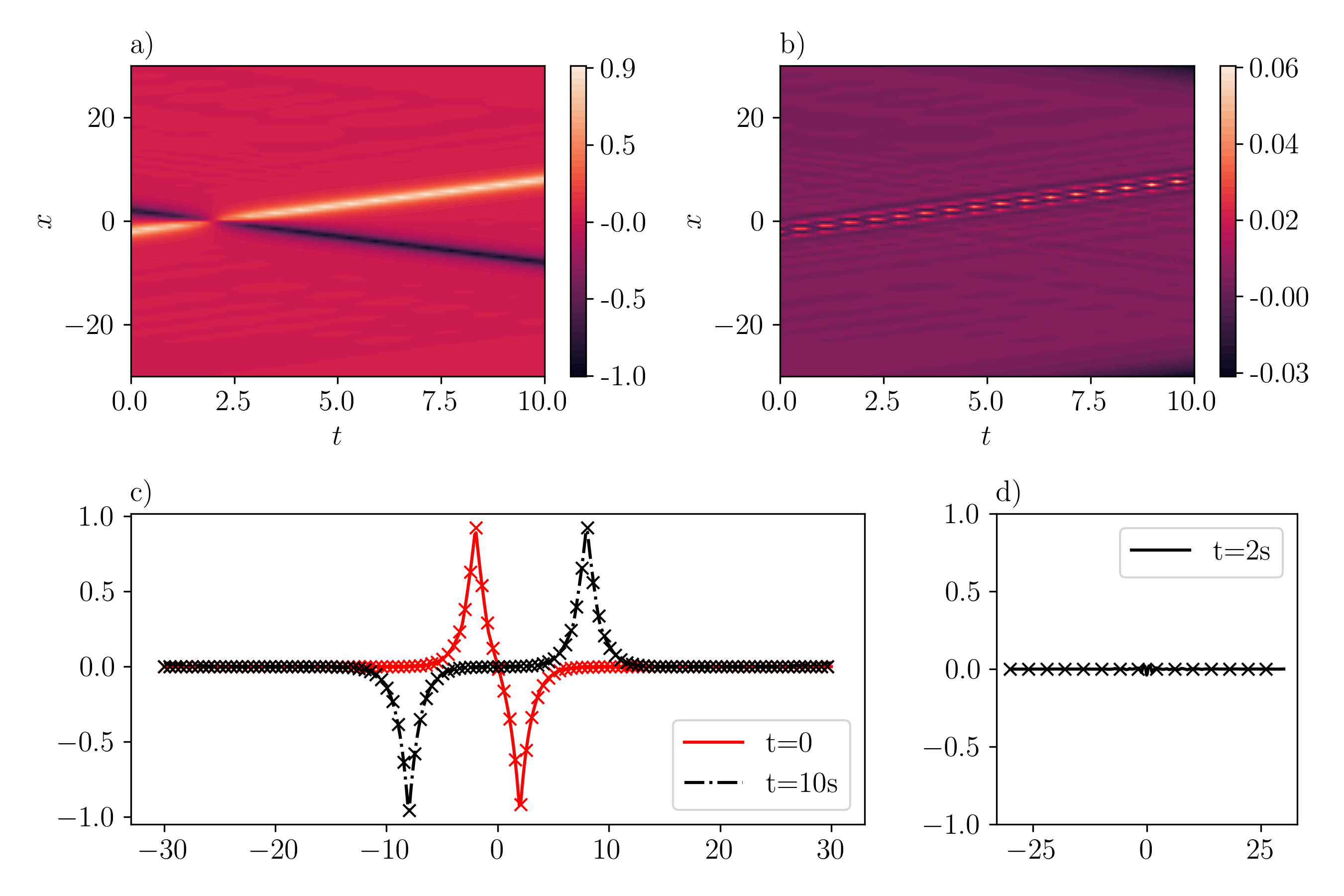}
    \caption{Same as Fig.~\ref{fig:fig7} but for the case corresponding to the
    "head-on" collision of a peakon and an anti-peakon (with $c_1=c_2=-1$)
    in the CH equation.~a): $\bar{u}(x,t)$ inferred by SGNN; b): error $e(x,t)=u(x,t)-\bar{u}(x,t)$;
    c): $\bar{u}(x,t)$ at two time instants.~The format of panel c) is the same as
    the one in Fig.~\ref{fig:fig6}.~The training loss is $6.12e-3$, with
    $\lambda_{ic}=\lambda_{bc}=100$.~Validation error: $\Vert e\Vert=5.54e-2$, $\Vert e\Vert_2=2.11e-5$. }

    \label{fig:fig8}
\end{figure}

\begin{figure}[!ht]
    \centering
    \includegraphics[width=0.9\textwidth]{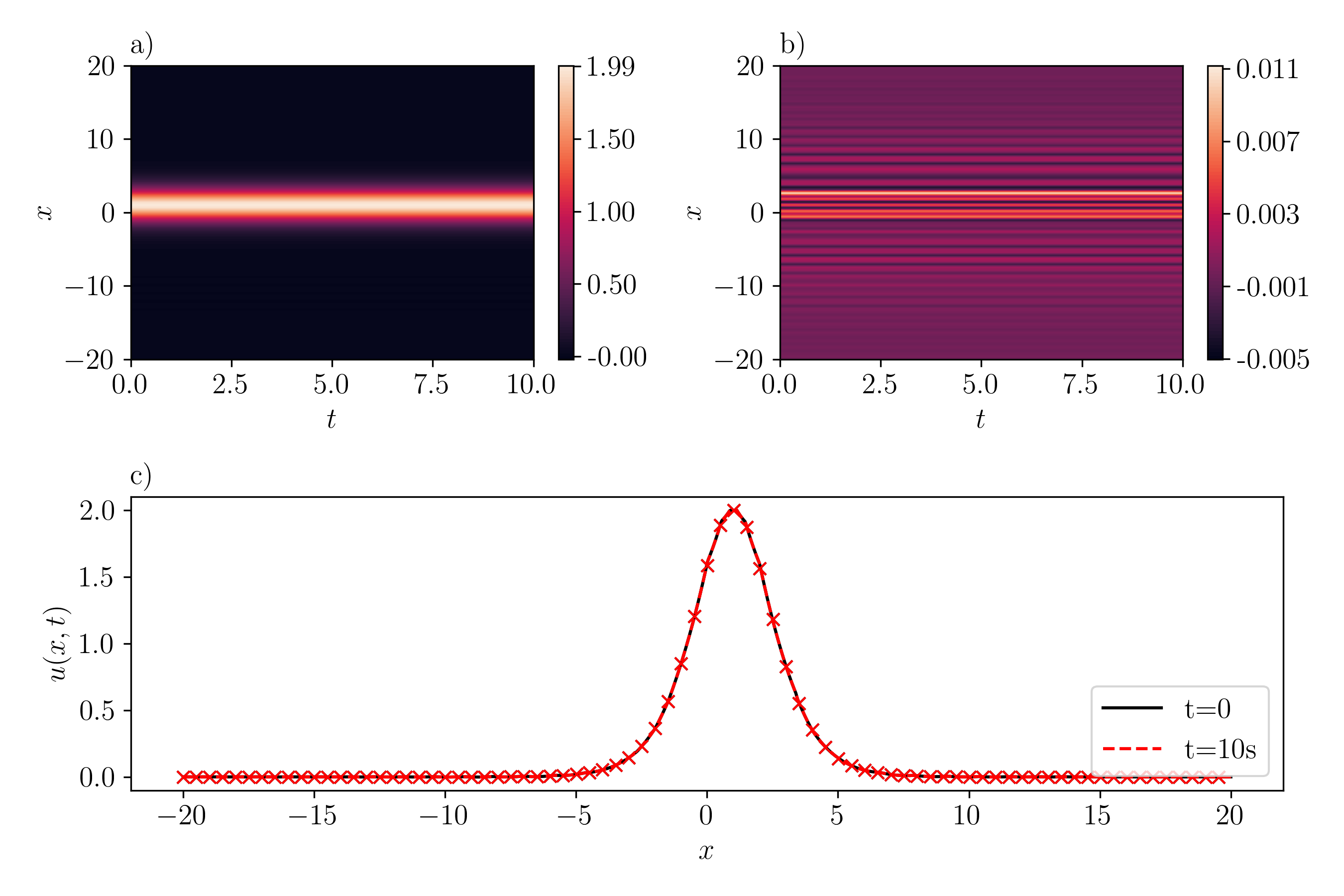}
    \caption{Same as Fig.~\ref{fig:fig8} but for a stationary solution, i.e., "lefton"
    of $b$-family with $b=-2.0$.~a): $\bar{u}(x,t)$ inferred by SGNN;
     b) $e(x,t)=u(x,t)-\bar{u}(x, t)$; c) $u(x,t)$ at two time instants.~%
     ~The format of panel c) is the same as the one in Fig.~\ref{fig:fig7}.%
     ~The training loss is 0.054, with $\lambda_{ic}=1000$, $\lambda_{bc}=1.$.%
     ~Validation loss: $\Vert e\Vert=1.12e-2$, $\Vert e\Vert_2=4.62e-6$.}
    \label{fig:fig9}
\end{figure}

The complementary case corresponding to the interaction of two anti-peakons
traveling at different speeds is presented in Fig.~\ref{fig:fig7}.~In particular,
we consider a configuration involving two anti-peakons: one centered at $x=-5$ with
speed $0.8$, and another one whose center is (symmetrically) placed at $x=5$ and
travels with velocity of $2.2$.~The respective IC that describes this configuration
is $u(x,0)=-0.8e^{-|x+5|}-2.2e^{-|x-5|}$.~The training error is $0.0188$, with
scaling factors $\lambda_{ic}=\lambda_{bc}=100$.~On the validation dataset, the
mean-squared error is $2.99e-5$.~In addition, the maximum absolute error is $5.22e-2$,
which is reflected in Fig.~\ref{fig:fig6}(b).~The interactions of these two anti-peakons
is shown in Fig. \ref{fig:fig7}(a).~The second anti-peakon (in darker red), possessing
a higher velocity, will eventually overtake the first one (in orange), despite initially
lagging behind.~A good agreement between SGNN prediction and exact solution is demonstrated
in Fig.~\ref{fig:fig7} (c).~The second one catches the first one at $t=7.14$, where their
peaks add up, as shown in Fig.~\ref{fig:fig7} (d).

Finally, Fig.~\ref{fig:fig8} shows a more realistic scenario: the (elastic) collision between
a peakon and an anti-peakon.~In this case, the IC considered is given by $u(x,0)=-e^{-|x-2|}+e^{-|x+2|}$,
where the training error is $6.12e-3$, with scaling factors $\lambda_{ic}=\lambda_{bc}=100$.%
~The mean-squared validation error is $2.11e-5$ while the maximum absolute validation error
is $5.54e-2$, as illustrated in Fig.~\ref{fig:fig8} (b).~As expected, the peakon (light red)
and anti-peakon (in darker red) move towards each other with same velocity as shown in Fig.~\ref{fig:fig8}(a)
until they collide at $t=2$.~Indeed, Fig.~\ref{fig:fig8}(d) showcases the predicted solution
at the time of collision where the waveforms cancel each other.~Then, at later times, i.e.,
$t>2$, the anti-peakon and peakon re-emerge, and they can maintain their shape after collision,
as shown in Fig. \ref{fig:fig8}(c).

\subsubsection{Lefton solutions}

The last case that we consider using SGNNs is the lefton regime, i.e., $b<-1$
whose explicit solution form is given by Eq.~\eqref{eq:leftons}.~Herein, we study such
solutions at $b=-2$.~For our training dataset, we randomly selected $2^{12}=4096$
points within the domain alongside an additional $2^9=512$ points, subsequently
dividing this dataset into 8 mini-batches.~The chosen time domain is set at
$t\in[-10,10]$, and the spatial domain at $x\in[-20,20]$.~As depicted in Figure~\ref{fig:fig9},
there is a high degree of concordance between the SGNN predictions and the exact
solutions.~The training loss, adjusted by scaling factors $\lambda_{ic}=1,000$
and $\lambda_{bc}=1$, was recorded at $0.054$. The mean-squared error for the
validation loss stands at $4.62e-6$, with the maximum absolute validation loss
reaching $1.12e-2$.

\section{Peakons in ab-family}

In this section, we turn our focus on the applicability of SGNN to the
so-called $ab$-family~\cite{Alexandrou-Himonas-ab-family-2016}
\begin{equation}
    u_t+u^2u_x-au_x^{3}+D^{-2}\partial_x\left[\frac{b}{3}u^3+\frac{6-6a-b}{2}uu_x^2\right] +D^{-2}\left[\frac{2a+b-2}{2}u_x^3\right]=0
    \label{eq:abfam}
\end{equation}
of peakon equations where $D^{-2}$ is a nonlocal operator in the form
$(1-\partial_x^2)^{-1}$.~The $ab$-family is a generalization of the
$b$-family [cf.~Eq.~\eqref{Eq:b-family}] in the sense that it corresponds
to cubic (in its nonlinearity) CH-type equations unlike the quadratic
CH-type equations of the $b$-family~\cite{Alexandrou-Himonas-ab-family-2016}.%
~Interestingly, the $ab$-family admits the one-peakon solution taking the form
\begin{equation}
    u(x,t)=\pm\sqrt{c}e^{-|x-(1-a)ct|}.
\end{equation}

For the applicability of SGNN, we inspect the peakon solution in the spatial
domain $x\in [-20, 20]$ and time domain $t\in [0, 10]$.~An SGNN with $80$ neurons
is used to approximate the one-peakon solution in the $ab$-family.~To generate
the training dataset, we randomly generate $2^{13}=8192$ collection samples
within the domain and $2^{10}=1024$ samples on the boundary.~The training dataset
is evenly split into $8$ mini-batches.~In the loss function, $\lambda_{ic}=1000$
and $\lambda_{bc}=100$ are applied to penalize initial and boundary conditions.%
~Same as before, the ADAM method is then used to train the SGNN, followed by the
refinement by L-BFGS.

\begin{figure}[!h]
    \centering
    \includegraphics[width=0.9\textwidth]{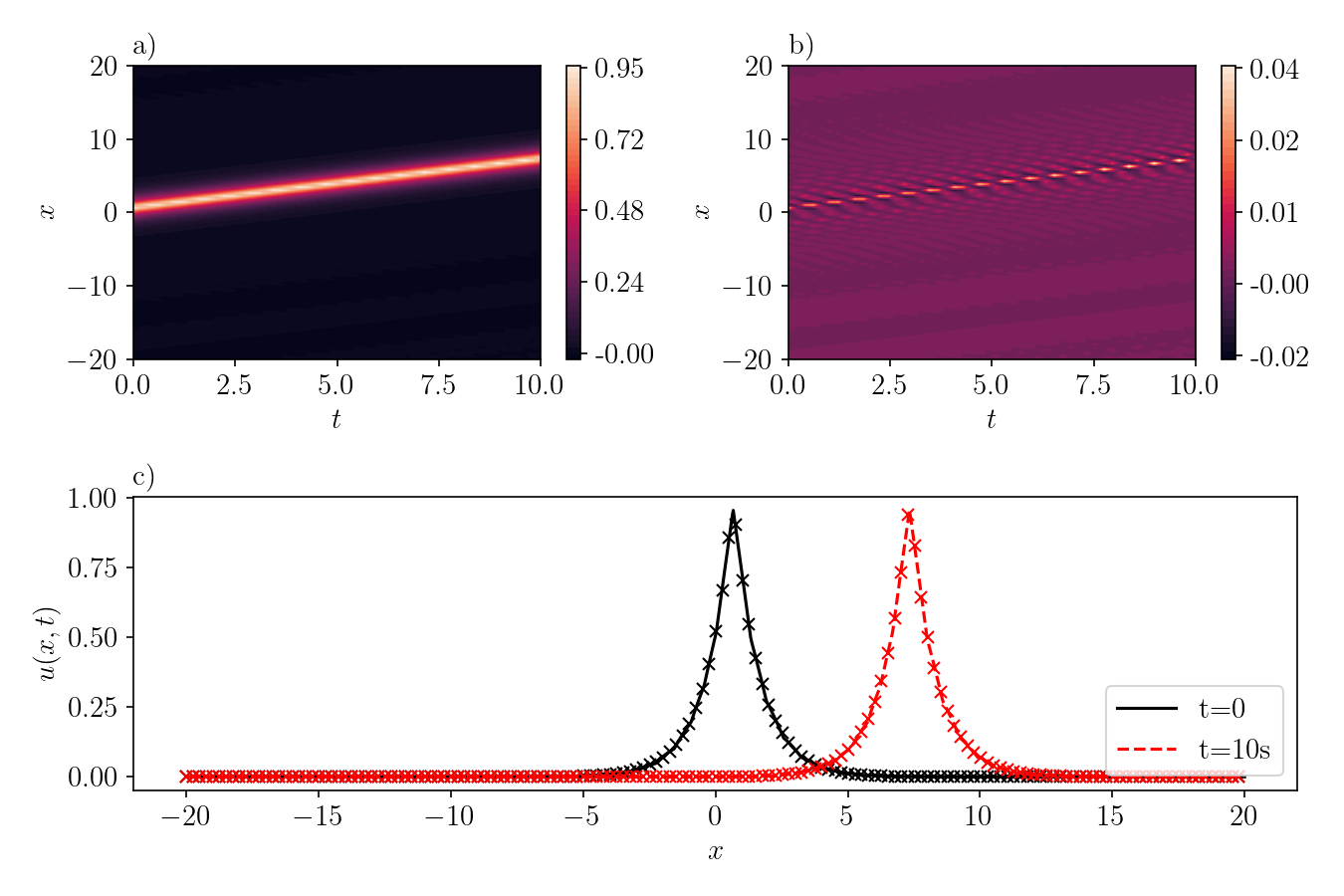}
    \caption{A peakon in the $ab$-family with $b=2.0, a=1/3$.~The wave speed is
     $c=1$.  a): $\bar{u}(x,t)$ inferred by SGNN; b) $e(x,t)=u(x,t)-\bar{u}(x, t)$;
     c) $u(x,t)$ at two time instants.~In c), "x" markers represent the exact solution
     while lines depict the prediction by SGNN.~The training loss is 0.0141, with
     $\lambda_{ic}=1000$, $\lambda_{bc}=100$. Validation loss: $\Vert e\Vert=3.70e-2$,
     $\Vert e\Vert_2=8.63e-6$. }
    \label{fig:fig10}
\end{figure}
\begin{figure}[!ht]
    \centering
    \includegraphics[width=0.9\textwidth]{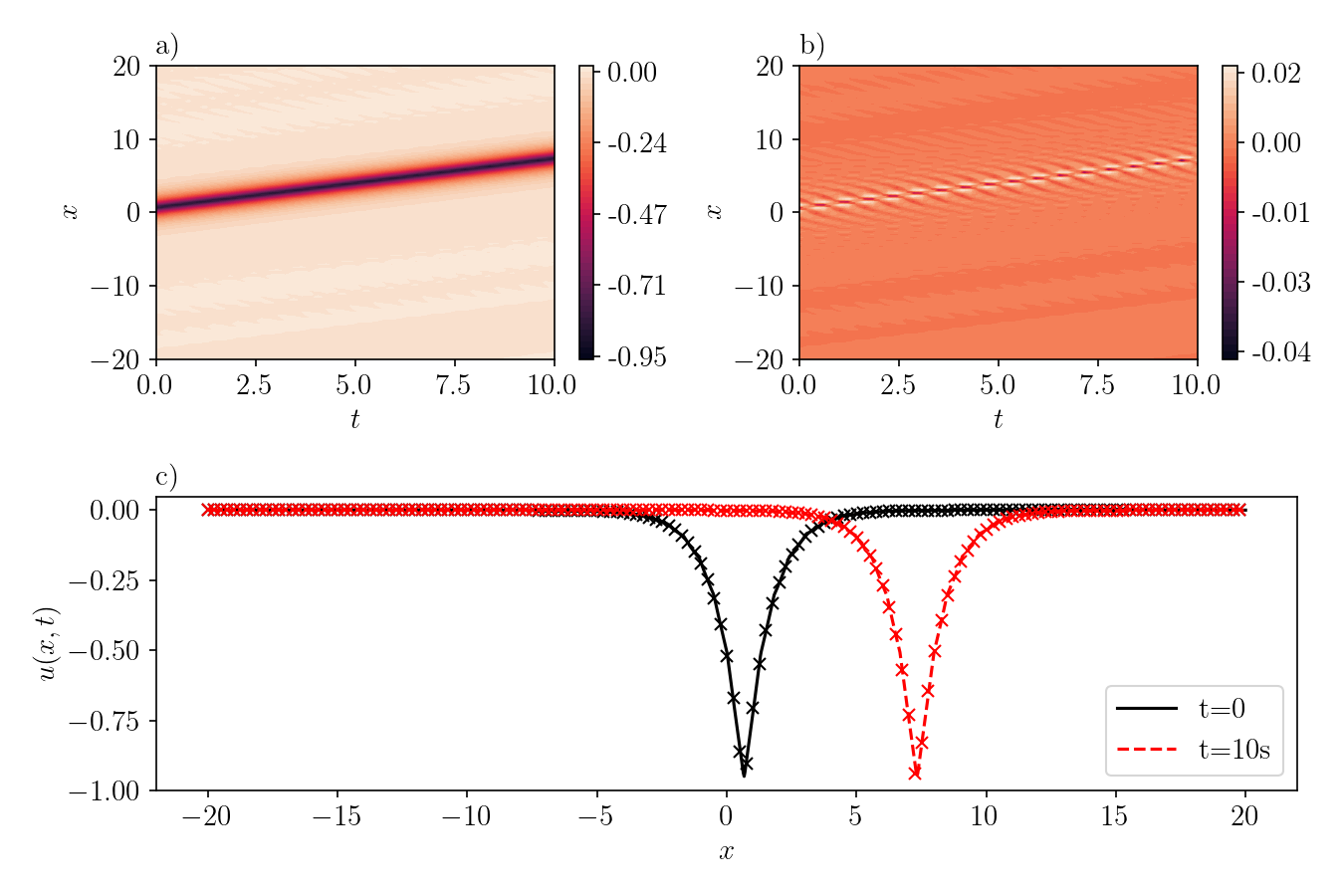}
    \caption{Same as in Fig.~\ref{fig:fig10}, but for an anti-peakon in the $ab$-family
    with $b=2.0, a=1/3$, and wave speed $c=-1$. a): $\bar{u}(x,t)$ inferred by SGNN;
    b) $e(x,t)=u(x,t)-\bar{u}(x, t)$; c) $u(x,t)$ at two time instants.~The format of
    panel c) is the same as the one in Fig.~\ref{fig:fig10}.~Here, the training loss
    is $0.0138$, with  $\lambda_{ic}=1000$, $\lambda_{bc}=100$.~Validation error:
    $\Vert e\Vert =4.05e-2$, $\Vert e\Vert _2=8.24e-6$.}
    \label{fig:fig11}
\end{figure}

Distinct from the members of the $b$-family, both peakons and anti-peakons of the
$ab$-family propagate in the same direction.~This behavior is confirmed using parameters
$b=2.0$, $a=1/3$, and $c=1$, as illustrated in Figs.~\ref{fig:fig10} and~\ref{fig:fig11}.%
~The training losses for the peakon and anti-peakon solutions are recorded at $0.0141$
and $0.0138$, respectively.~For the peakon solution, the mean-squared error across the
validation set is measured at $8.63e-6$, with the maximum absolute error reaching $3.7e-2$.%
~Similarly, the anti-peakon solution exhibits a mean-squared error of $8.24e-6$ over the
validation set, and its maximum absolute error is noted as $4.5e-2$.

\section{2D compactons}
In this section, we depart from the previous one-dimensional (in space) studies, and
apply SGNN in order to predict TWs in two-dimensional nonlinear wave equations.%
~More specifically, we focus on TWs that have compact support which are referred
to as compactons, and introduced in~\cite{Rosenau2000} (and references therein).%
~Following the notation in~\cite{Rosenau2000}, there exists a family of PDEs
denoted as $C_N(m,a+b)$ given by
\begin{equation}
    u_t + (u^m)_x + \frac{1}{b}[u^a(\triangledown^2u^b)]_x=0,
    \label{eq: 2D Compactons.}
\end{equation}
that possesses compacton TWs with $m\geq max(1, a-1)$, $b>0$.~Here, $C_N(m,a+b)$
represents a $N$-dimensional compacton (with $N=1,2,3$) with a parameter set
\{$m$, $a$, $b$\}.~In the following, we restrict ourselves to $N=2$.%
~According to \cite{Rosenau_compacton_2007}, Eq.~\ref{eq: 2D Compactons.} supports
traveling compactons traversing in the $x$ direction.~In that light, the
characteristic is in the form
\begin{equation}
    s=x-\lambda t,
\end{equation}
where $\lambda$ is the velocity of the compacton.~The case with $C_2(m=1+b, 1+b)$
yields an explicit solution in the form
\begin{equation}
    u=\lambda^{1/b}\left[1-\frac{F(R)}{F(R^*)}\right]^{1/b},~ 0<R\leq R^*,
    \label{eq:2D Compacton-(3,1+2)}
\end{equation}
where $u$ vanishes elsewhere (i.e., compact support).~In Eq.~\eqref{eq:2D Compacton-(3,1+2)},
$R=\sqrt{s^2+y^2}$, and $F(R)=J_0(\sqrt{b}R)$ where $J_{0}$ is the zeroth-order
Bessel function, and $\sqrt{b}R^*$ is the root of the first-order Bessel function.

\begin{figure}[!pt]
    \centering
    \includegraphics[width=0.9\textwidth]{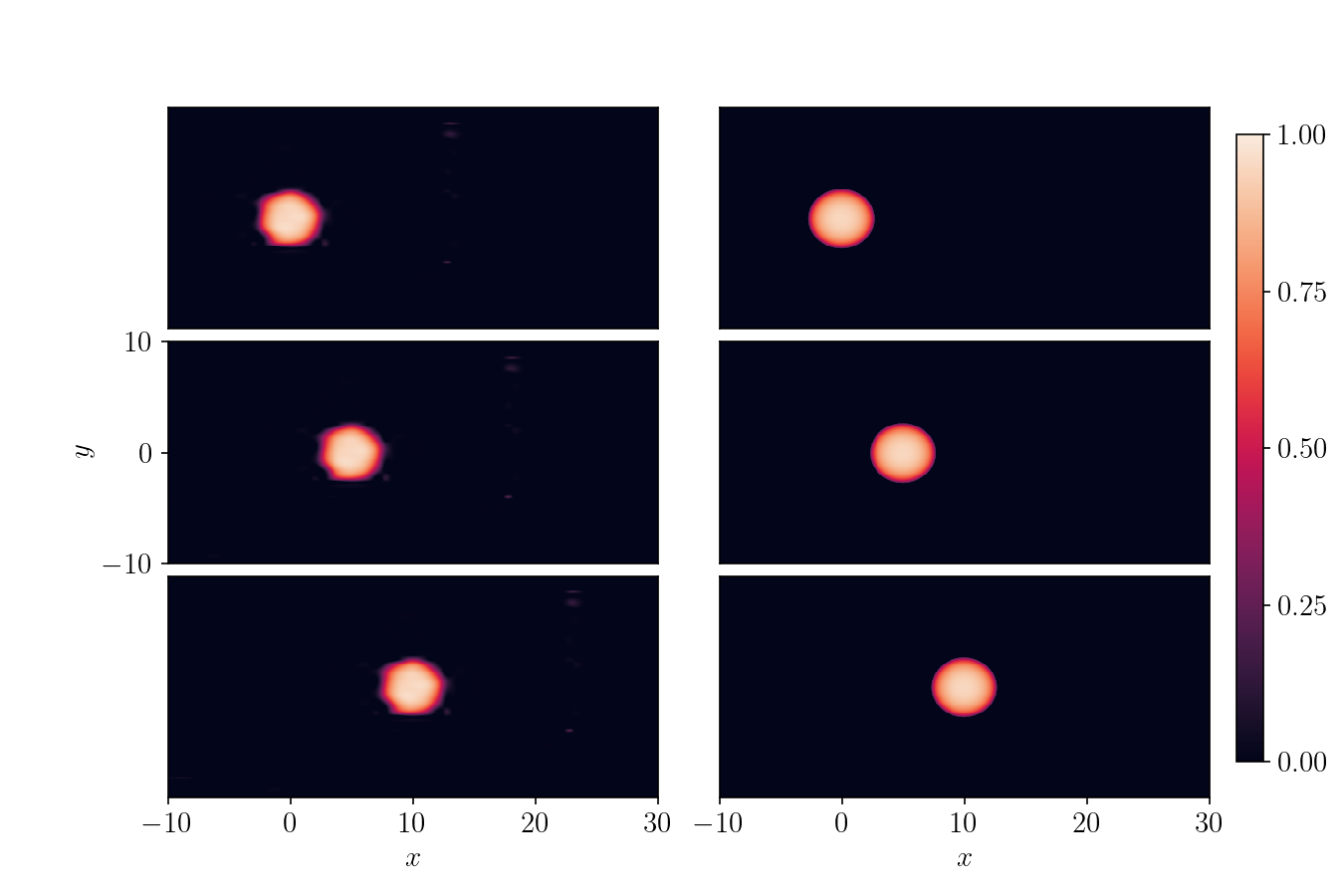}
     \caption{A 2D compacton $C_2(3,1+2)$ of Eq.~\eqref{eq:2D Compacton-(3,1+2)}
     with $\lambda=1$.~Left panel: SGNN prediction; right panel: ground truth.%
     ~Top panel: $t=0$; middle panel: $t=5$; bottom panel: $t=10$. The training loss:$9.97e-3$:$\lambda_{ic}=100$, $\lambda_{bc}=100$. Validation error:
     $\Vert e\Vert =0.371$, $\Vert e\Vert _2=1.58e-4$.}
    \label{fig:fig12}
\end{figure}

\begin{figure}[!ht]
    \centering
    \includegraphics[width=0.9\textwidth]{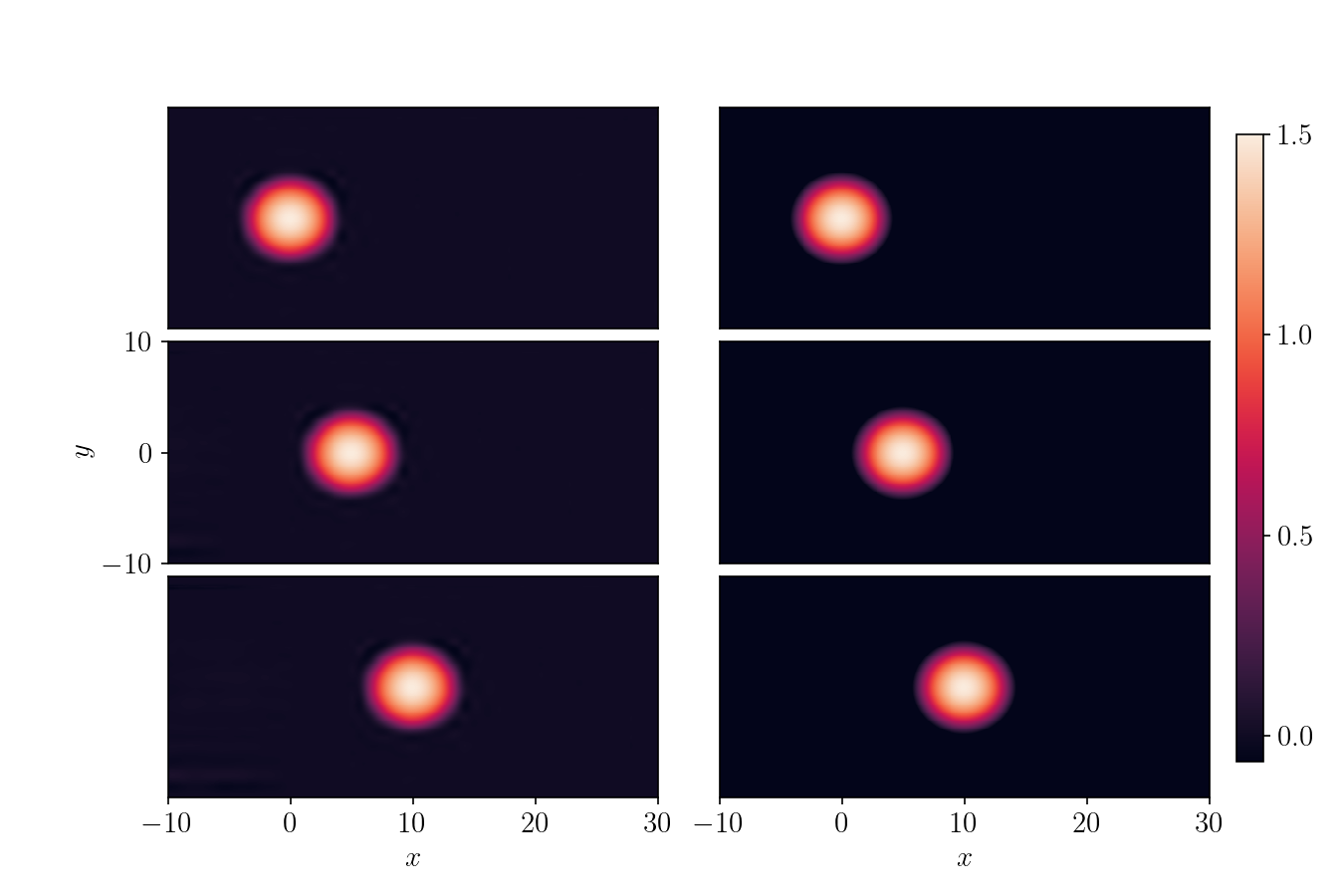}
     \caption{Same as Fig.~\ref{fig:fig12} but for the 2D compacton $C_2(2,0+3)$
     of Eq.~\eqref{eq:2D COmpacton-(2,0+3)} with $\lambda=1$.%
     ~Left panel: SGNN prediction; right panel: ground truth.%
     ~Top panel: $t=0$; middle panel $t=5$, bottom panel: $t=10$.The training loss:$1.23e-3$:$\lambda_{ic}=100$, $\lambda_{bc}=10$. Validation error:
     $\Vert e\Vert =0.135$, $\Vert e\Vert _2=8.28e-5$.}
    \label{fig:fig13}
\end{figure}

We use a SGNN to approximate the compacton $C_2(3,1+2)$.~The SGNN has two layers,
with $50$ neurons per layer, and the approximation is performed in the spatial
domain $x\in[-10,30]$ and time domain $t\in[0,10]$.~To generate the training dataset,
we randomly sampled $ 2^{16}=65536$ collocation points within the domain, along with
$2^{12}=4096$ points on the boundary.~The dataset is then evenly split into $8$
mini-batches.~The mini-batch ADAM is used to SGNN, with loss functions to minimize
the residual error of the PDE, initial conditions, and boundary conditions.

In Fig.~\ref{fig:fig12}, the SGNN's prediction (left column) is presented against the exact
(right column) compacton solution $C_2(3,1+2)$.~The training loss is stopped at $9.97e-3$,
with $\lambda_{ic}=100$, and $\lambda_{bc}=10$.~The mean-squared validation error is $1.58e-4$
while the maximum absolute error is $0.371$.~The left panel is predicted by SGNN and the right
panel shows the exact solution.~The compacton travels along $x$-axis with a velocity of $\lambda=1$.%
~At $t=0$ (top panel), the compacton commences with a center placed at $x=0$.~Middle and
bottom panels present snapshots of compactons at $t=5$, and $t=10$, respectively.~We report
that the SGNN's prediction captures the main characteristics of the exact solution although minor
errors appear around the edges of compacton.

As a last case, we consider the compacton $C_2(m=2,a+b=3)$ whose explicit solution
is given by
\begin{equation}
    u=\kappa_N[\lambda A_N - bR^2], 0<R\leq R_*\equiv\sqrt{\lambda A_N/b}
\end{equation}
with $u$ vanishing elsewhere.~According to Ref.~\cite{Rosenau_compacton_2007}, we pick
$N=2$, $m=0$, $a=0$, and $b=3$, and thus we have
\begin{equation}
    C_N(2,0+3): A_2=\frac{3}{2}(4+2)^2, \kappa_2^{-1}=6(4+2).
    \label{eq:2D COmpacton-(2,0+3)}
\end{equation}
The SGNN's prediction and exact solution of $C_2(2,0+3)$ compacton solutions are presented
in the left and right columns of Fig.~\ref{fig:fig13}.~Snapshots of the solutions are shown
at $t=0$ (top), $t=5$ (middle), and $t=10$ (bottom) therein.~In this case, we report that
the predictions precisely match the exact solution at these times.~The scaling factors in
the loss function are $\lambda_{ic}=100$ and $\lambda_{bc}=10$, and the training loss is
stopped at $1.23e-3$ after $300$ epochs.~The mean-squared validation error is $8.28e-5$,
while the maximum absolute error is $0.135$.

In summary, we present the training losses and validation errors of all previous results
(see, also the reference Figure in the left column) in Table~\ref{table:losses}.


\begin{table}[!ht]
\centering
\caption{The training losses and validation errors of the presented results in sections 2-5.}
\begin{tabular}{cccc}
\toprule
\multirow{2}{*}{Figure} & \multirow{2}{*}{Training loss} & \multicolumn{2}{c}{Validation error } \\
\cmidrule(lr){3-4}
& &   $\Vert e \Vert_2$                 &  $\Vert e \Vert$ \\ 

\hline
3      & $8.43e-3$     & $7.21e-6$                                            & $3.90e-2$                        \\
4      & $1.94e-11$    & $9.59e-13$                                           & $1.02e-5$                        \\
5      & $9.18e-2$     & $4.09e-6$                                            & $2.74e-2$                        \\
6      & $3.10e-2$     & $3.50e-5$ & $5.92e-3$                        \\
8      & $4.27e-3$     & $2.99e-5$                                            & $5.22e-2$                        \\
9      & $1.88e-2$     & $9.12e-5$                                            & $9.00e-2$                        \\
10     & $6.12e-3$     & $2.11e-5$                                            & $5.54e-2$                        \\
11     & $5.40e-2$       & $4.62e-6$                                            & $1.12e-2$                        \\
12     & $1.40e-2$       & $8.63e-6$                                            & $3.70e-2$                        \\
13     & $1.38e-2$      & $8.24e-6$                                            & $4.05e-2$                        \\
14     & $9.97e-3$     & $1.58e-4$                                            & $3.71e-1$                          \\
15     & $1.23e-3$     & $8.28e-5$                                             & $1.35e-1$                          \\ \bottomrule
\end{tabular}
\label{table:losses}
\end{table}

\section{Comparison and discussion}
To compare the performance of the traditional and new structures of PINNs, we use them
to approximate a peakon solution in the CH equation with $c=1$.~The spatial domain employed
is $[-20, 20]$, while the temporal domain is $[0,10]$.~The size of the training set is $2^{12}=4096$,
with $2^9=512$ samples at the initial and boundary conditions.~The selection of width and depth
for models is informed by the configurations reported in existing literature.~Additionally, we
also compare the performance of SGNN vs MLP.~As shown in Table~\ref{table: 1}, in the traditional
PINN framework, neither SGNN nor MLP can successfully converge to a TW on a large spatial and
temporal domain.~Despite small training losses, all NN structures get stuck at the trivial solution
as it is very difficult to overcome propagation failure when dealing with enlarged domains.~By
introducing a training method that respects causality~\cite{wang2024respectingcausality} or
performs adaptive sampling~\cite{WuCMAME2023}, one may be able to address this failure.

On the other hand, as illustrated in Table~\ref{table: 2}, with the TW coordinate transformation,
identical NN structures of SGNN and MLP with ReLu function all converge to the correct solution.%
~Although the training losses of MLP with hyperbolic tangent and sigmoid functions are relatively
large, they all capture the characteristics of the TW.~Sigmoid and hyperbolic functions have difficulties
approximating the non-differentiable peak of the peakon, with about $0.3$ error in the peak value.%
~Notably, these results can be improved by modifying the sampling method, training scheme, and loss
functions.~With the increase of depth and width, MLP with ReLU and sigmoid functions can further reduce
loss values.~The loss values with SGNN also gradually drop as width increases. SGNN excels at the compact
structure that only requires less than $1/10$ of training parameters.~In addition, SGNN can give an explicit
solution form of the PDEs in the sense of Gaussian radial-basis functions.~However, further increasing the
number of neurons in SGNN does not appreciably reduce the loss value.~This could be remedied by modifying
the training and sampling schemes.

\begin{table}[!ht]
\caption{Comparison of SGNN and MLP with the traditional PINNs. Despite small loss values, no network structure can converge to the correct solution. Spatial domain: [-30,30], time domain: [0, 10]. GRB: generalized Gaussian radial-basis function.}
\begin{tabular}{l|lcccc}
\toprule
Network               & Activation & Depth & Width & Loss              & Trivial solution? \\ \hline
SGNN                  & GRBF        & 1     & 40    & $(2.69\pm5.47)e-7$   & Yes               \\\hline
\multirow{12}{*}{MLP} & ReLu       & 2     & 40    & $(1.13\pm0.60)e-3$   & Yes               \\
                      & ReLu       & 4     & 40    &  $(6.11\pm4.10)e-4$            & Yes               \\
                      & ReLu       & 6     & 40    & $(1.05\pm0.66)e-3$ & Yes               \\
                      & ReLu       & 8     & 20    &  $(1.06\pm0.77)e-3$                  & Yes               \\
                      & sigmoid    & 2     & 40    &  $(7.37\pm2.10)e-6$                 & Yes               \\
                      & sigmoid    & 4     & 40    &   $(1.28\pm0.62)e-5$                & Yes               \\
                      & sigmoid    & 6     & 40    & $(1.08\pm0.82)e-6$   & Yes               \\
                      & sigmoid    & 8     & 20    & $(1.91\pm0.91)e-5$                  & Yes               \\
                      & tanh       & 2     & 40    &  $(1.26\pm0.24)e-6$                 & Yes               \\
                      & tanh       & 4     & 40    &   $(2.11\pm 0.54)e-6$                & Yes               \\
                      & tanh       & 6     & 40    & $(2.24\pm0.69)e-6$   & Yes               \\
                      & tanh       & 8     & 20    & $(3.59\pm1.46)e-6$                  & Yes               \\ \bottomrule
\end{tabular}
\label{table: 1}
\end{table}

\begin{table}[!ht]
\caption{Comparison of SGNN and MLP with PINNs incorporating with traveling frame. Spatial domain: [-30,30], time domain: [0, 10]. GRB: generalized Gaussian radial-basis function.}
\begin{tabular}{l|lccccc}
\toprule
Network               & Activation & Depth & Width &Parameters & Loss              & Trivial  \\
&&&&&& solution?\\
\hline
                  & GRBF        & 1     & 20 & 60  & $(1.42\pm0.08)e-2$    & No               \\
SGNN                  & GRBF        & 1     & 40 & 120  & $(1.02\pm0.17)e-2$   & No               \\
                  & GRBF        & 1     & 60 & 180  &  $(8.65\pm1.22)e-3$ & No               \\\hline
\multirow{12}{*}{MLP} & ReLu       & 2     & 40 & 1640  & $(2.95\pm0.12)e-2$   & No               \\
                      & ReLu       & 4     & 40 &  4840  &  $(2.93\pm0.11)e-2$                 & No               \\
                      & ReLu       & 6     & 40 &  8040 & $(5.34\pm2.10)e-4$ & No               \\
                      & ReLu       & 8     & 20 & 2820  &    $(8.60\pm2.93)e-4$               & No               \\
                      & sigmoid    & 2     & 40 & 1640  &  $0.72\pm5.62e-4$                 & No               \\
                      & sigmoid    & 4     & 40 & 4840  &    $0.72\pm 6.80e-3$               & No               \\
                      & sigmoid    & 6     & 40 &  8040 &  $0.71\pm7.61e-3$  & No               \\
                      & sigmoid    & 8     & 20 & 2820  &  $0.72\pm5.33e-3$                & No               \\
                      & tanh       & 2     & 40 & 1640  &  $0.72\pm7.00e-3$                 & No               \\
                      & tanh       & 4     & 40 & 4840  &  $0.71\pm4.65e-3$                 & No               \\
                      & tanh       & 6     & 40 & 8040  & $0.70\pm0.05$   & No               \\
                      & tanh       & 8     & 20 & 2820  &  $0.56\pm0.15$                 & No               \\ \bottomrule
\end{tabular}
\label{table: 2}
\end{table}
Why can the modified structure of PINNs avoid propagation failure and lead to better results? We attempt
to answer this question next.~By mathematically transforming the NN input to the traveling coordinate $x-ct$,
we inherently produce an output in the form of $u(x-ct)$.~This representation naturally aligns with the
solution form of TWs, maintaining the integrity of the solution's structure.~From a physical perspective,
this transformation converts a dynamic problem into a static one (i.e., a TW becomes stationary in a frame that
co-moves with the solution), thus simplifying the problem considerably.~Algorithmically, this transformation
effectively reduces the input dimension by one, which can lead to a decrease in the required data size for
training.~Furthermore, the functional form of TWs of $u(x-ct)$ ensures that any combination of spatial and
temporal coordinates resulting in the same traveling frame coordinate will produce identical outcomes.%
~This characteristic automatically propagates initial and/or boundary conditions along the characteristic
path $x-ct$, significantly reducing the challenge of extending solutions from initial and boundary conditions
to interior points.

\section{Conclusions}

In this work, we introduce a modified structure of PINNs that incorporates the mathematical solution
form of TWs to nonlinear PDEs.~In particular, we integrated a novel neural network architecture, called
SGNN, into the PINNs framework.~Our approach demonstrated a significant improvement in overcoming the
propagation failure of PINNs, particularly in large-domain applications.~Utilizing this enhanced network,
we successfully generated interpretable predictions for TWs across various PDE families including the $b$-
and $ab$-families of peakon equations.~To the authors' best knowledge, this is also the first study on
applying PINNs to identify 2D TWs, such as compactons, as well as to study the collisions of 1D multiple
peakons.~This work opens up new directions for future studies that we plan to undertake herein.~Specifically,
and on the one hand, there exist solutions to nonlinear dispersive PDEs that self-similarly blow-up in finite
(or infinite) time~\cite{sulem_book}.~Under a stretching transformation~\cite{sulem_book}, such solutions
can appear as steady ones in a frame that "co-explodes" with the solution~\cite{CHAPMAN2022133396,kdvchapman},
thus enabling the applicability of the present NN architecture for the identification and prediction of self-similar
collapse.~On the other hand, the present NN structure could be expanded in order to model and predict the
transient behavior of TWs.~In addition, regularization techniques can be incorporated to refine the model to
capture the essential features of the solutions more succinctly.

\vspace{6pt} 




\authorcontributions{Conceptualization, methodology, software, validation,
writing---original draft preparation: S. Xing.~Suggesting problems,
writing---review and editing: E.G.~Charalampidis.~All authors have
read and agreed to the published version of the manuscript.}

\funding{S. Xing is supported by the Donald E. Bently Center for Engineering
Innovation and Lockheed Endowed Professorship in the College of Engineering
at Cal Poly.~This work has been supported by the U.S. National
Science Foundation under Grant no. DMS-2204782 (EGC).}

\dedication{We dedicate this work to the memory of our esteemed colleague and friend,
Professor Joseph Callenes-Sloan (Electrical and Computer Engineering Department,
Cal Poly San Luis Obispo).~}




\begin{adjustwidth}{-\extralength}{0cm}

\reftitle{References}


\bibliographystyle{unsrt}
\bibliography{cas-refs}

\begin{thebibliography}{999}

\bibitem[Raissi et~al.(2019)Raissi, Perdikaris, and
  Karniadakis]{RaissiPINN2019}
Raissi, M.; Perdikaris, P.; Karniadakis, G.
\newblock Physics-informed neural networks: A deep learning framework for
  solving forward and inverse problems involving nonlinear partial differential
  equations.
\newblock {\em Journal of Computational Physics} {\bf 2019}, {\em
  378},~686--707.

\bibitem[Karniadakis et~al.(2021)Karniadakis, Kevrekidis, Lu, Perdikaris, Wang,
  and Yang]{KarniadakisNature2021}
Karniadakis, G.; Kevrekidis, I.; Lu, L.; Perdikaris, P.; Wang, S.; Yang, L.
\newblock Physics-informed machine learning.
\newblock {\em Nature Reviews Physics} {\bf 2021}, {\em 3},~422--440.

\bibitem[Cai et~al.(2021)Cai, Wang, Wang, Perdikaris, and
  Karniadakis]{CaiJHT2021}
Cai, S.; Wang, Z.; Wang, S.; Perdikaris, P.; Karniadakis, G.
\newblock {Physics-informed neural networks for heat transfer Problems}.
\newblock {\em Journal of Heat Transfer} {\bf 2021}, {\em 143},~060801.

\bibitem[Jin et~al.(2021)Jin, Cai, Li, and Karniadakis]{JinJCP2021}
Jin, X.; Cai, S.; Li, H.; Karniadakis, G.
\newblock NSFnets (Navier-Stokes flow nets): Physics-informed neural networks
  for the incompressible Navier-Stokes equations.
\newblock {\em Journal of Computational Physics} {\bf 2021}, {\em 426},~109951.

\bibitem[Wang et~al.(2023)Wang, Lai, G\'omez-Serrano, and
  Buckmaster]{PhysRevLett.130.244002}
Wang, Y.; Lai, C.Y.; G\'omez-Serrano, J.; Buckmaster, T.
\newblock Asymptotic self-similar blow-up profile for three-dimensional
  axisymmetric euler equations Using Neural Networks.
\newblock {\em Phys. Rev. Lett.} {\bf 2023}, {\em 130},~244002.

\bibitem[Zhu et~al.(2022)Zhu, Khademi, Charalampidis, and
  Kevrekidis]{WeiZhuSPINN_2022}
Zhu, W.; Khademi, W.; Charalampidis, E.G.; Kevrekidis, P.G.
\newblock Neural networks enforcing physical symmetries in nonlinear dynamical
  lattices: The case example of the {A}blowitz–{L}adik model.
\newblock {\em Physica D} {\bf 2022}, {\em 434},~133264.

\bibitem[Saqlain et~al.()Saqlain, Zhu, Charalampidis, and
  Kevrekidis]{SAQLAIN2023107498}
Saqlain, S.; Zhu, W.; Charalampidis, E.G.; Kevrekidis, P.G.

\bibitem[Van~Herten et~al.(2022)Van~Herten, Chiribiri, Breeuwer, Veta, and
  Scannell]{vanHerten2022}
Van~Herten, R.L.; Chiribiri, A.; Breeuwer, M.; Veta, M.; Scannell, C.M.
\newblock Physics-informed neural networks for myocardial perfusion mri
  quantification.
\newblock {\em Medical Image Analysis} {\bf 2022}, {\em 78},~102399.

\bibitem[Wang et~al.(2021)Wang, Teng, and Perdikaris]{Wangeltl2021}
Wang, S.; Teng, Y.; Perdikaris, P.
\newblock Understanding and mitigating gradient flow pathologies in
  physics-informed neural networks.
\newblock {\em SIAM Journal on Scientific Computing} {\bf 2021}, {\em
  43},~A3055--A3081.

\bibitem[Wang et~al.(2022)Wang, Yu, and Perdikaris]{WangJCP2022}
Wang, S.; Yu, X.; Perdikaris, P.
\newblock When and why PINNs fail to train: A neural tangent kernel
  perspective.
\newblock {\em Journal of Computational Physics} {\bf 2022}, {\em 449},~110768.

\bibitem[Krishnapriyan et~al.(2021)Krishnapriyan, Gholami, Zhe, Kirby, and
  Mahoney]{krishnapriyan2021characterizing}
Krishnapriyan, A.S.; Gholami, A.; Zhe, S.; Kirby, R.; Mahoney, M.W.
\newblock Characterizing possible failure modes in physics-informed neural
  networks.
\newblock {\em Advances in Neural Information Processing Systems} {\bf 2021},
  {\em 34}.

\bibitem[Daw et~al.(2023)Daw, Bu, Wang, Perdikaris, and
  Karpatne]{daw2023mitigating}
Daw, A.; Bu, J.; Wang, S.; Perdikaris, P.; Karpatne, A.
\newblock Mitigating Propagation Failures in Physics-informed Neural Networks
  using Retain-Resample-Release (R3) Sampling.
\newblock In Proceedings of the ICML'23: Proceedings of the 40th International
  Conference on Machine Learning, Honolulu, HI,  July 2023; p. 7264–7302.

\bibitem[Camassa and Holm(1993)]{PhysRevLett.71.1661}
Camassa, R.; Holm, D.D.
\newblock An integrable shallow water equation with peaked solitons.
\newblock {\em Phys. Rev. Lett.} {\bf 1993}, {\em 71},~1661--1664.

\bibitem[Tancik et~al.(2020)Tancik, Srinivasan, Mildenhall, Fridovich-Keil,
  Raghavan, Singhal, Ramamoorthi, Barron, and Ng]{tancik2020fourfeat}
Tancik, M.; Srinivasan, P.P.; Mildenhall, B.; Fridovich-Keil, S.; Raghavan, N.;
  Singhal, U.; Ramamoorthi, R.; Barron, J.T.; Ng, R.
\newblock Fourier Features Let Networks Learn High Frequency Functions in Low
  Dimensional Domains.
\newblock {\em NeurIPS} {\bf 2020}.

\bibitem[Wu et~al.(2023)Wu, Zhu, Tan, Kartha, and Lu]{WuCMAME2023}
Wu, C.; Zhu, M.; Tan, Q.; Kartha, Y.; Lu, L.
\newblock A comprehensive study of non-adaptive and residual-based adaptive
  sampling for physics-informed neural networks.
\newblock {\em Computer Methods in Applied Mechanics and Engineering} {\bf
  2023}, {\em 403(A)},~115671.

\bibitem[Wang et~al.(2024)Wang, Sankaran, and
  Perdikaris]{wang2024respectingcausality}
Wang, S.; Sankaran, S.; Perdikaris, P.
\newblock Respecting causality is all you need for training physics-informed
  neural networks.
\newblock {\em Computer Methods in Applied Mechanics and Engineering} {\bf
  2024}, {\em 421},~116813.

\bibitem[Wang and Yan(2021)]{WangandYanPhysicaD2021}
Wang, L.; Yan, Z.
\newblock Data-driven peakon and periodic peakon solutions and parameter
  discovery of some nonlinear dispersive equations via deep learning.
\newblock {\em Physica D} {\bf 2021}, {\em 428},~133037 (15 pages).

\bibitem[Braga-Neto()]{CINN_Barga_Neto_2023}
Braga-Neto, U.
\newblock Characteristics-informed neural networks for forward and inverse
  hyperbolic problems.
\newblock  \href{http://xxx.lanl.gov/abs/2212.14012}{{\normalfont
  [arXiv:quant-ph/2212.14012]}}.

\bibitem[Holm and Staley(2003)]{HOLM2003437}
Holm, D.D.; Staley, M.F.
\newblock Nonlinear balance and exchange of stability in dynamics of solitons,
  peakons, ramps/cliffs and leftons in a 1+1 nonlinear evolutionary PDE.
\newblock {\em Physics Letters A} {\bf 2003}, {\em 308},~437--444.

\bibitem[Himonas and Mantzavinos(2016)]{Alexandrou-Himonas-ab-family-2016}
Himonas, A.A.; Mantzavinos, D.
\newblock An ab-family of the equation with peakon traveling waves.
\newblock {\em Proceeding of the American Mathematical Society} {\bf 2016},
  {\em 144},~3797--3811.

\bibitem[Degasperis et~al.(2002)Degasperis, Holm, and Hone]{Degasperis2002}
Degasperis, A.; Holm, D.D.; Hone, A.N.W.
\newblock A new integrable equation with peakon solutions.
\newblock {\em Theoretical and Mathematical Physics} {\bf 2002}, {\em
  133},~1463--1474.

\bibitem[Fuchssteiner and Fokas(1981)]{FUCHSSTEINER198147}
Fuchssteiner, B.; Fokas, A.
\newblock Symplectic structures, their Bäcklund transformations and hereditary
  symmetries.
\newblock {\em Physica D: Nonlinear Phenomena} {\bf 1981}, {\em 4},~47--66.

\bibitem[Rosenau et~al.(2007)Rosenau, Hyman, and
  Staley]{Rosenau_compacton_2007}
Rosenau, P.; Hyman, J.M.; Staley, M.
\newblock Multidimensional Compactons.
\newblock {\em Physical Review Letters} {\bf 2007}, {\em 98},~024101.

\bibitem[Rosenau(2000)]{Rosenau2000}
Rosenau, P.
\newblock Compact and noncompact dispersive patterns.
\newblock {\em Physics Letters, Section A: General, Atomic and Solid State
  Physics} {\bf 2000}, {\em 275},~193--203.

\bibitem[Xing and Sun(2023)]{XingandSunSGNN2023}
Xing, S.; Sun, J.Q.
\newblock Separable Gaussian Neural Networks: Structure, analysis, and function
  approximations.
\newblock {\em Algorithms} {\bf 2023}, {\em 16},~453 (19 pages).

\bibitem[Park and Sandberg(1991)]{park_universal_1991}
Park, J.; Sandberg, I.W.
\newblock Universal approximation using Radial-Basis-Function Networks.
\newblock {\em Neural Computation} {\bf 1991}, {\em 3},~246--257.

\bibitem[Kingma and Ba(2015)]{DBLP:journals/corr/KingmaB14}
Kingma, D.P.; Ba, J.
\newblock Adam: {A} {M}ethod for {S}tochastic {O}ptimization.
\newblock In Proceedings of the ICLR (Poster),  2015.

\bibitem[Liu and Nocedal(1989)]{liu1989limited}
Liu, D.C.; Nocedal, J.
\newblock On the limited memory {BFGS} method for large scale optimization.
\newblock {\em Mathematical programming} {\bf 1989}, {\em 45},~503--528.

\bibitem[Kodama(1985{\natexlab{a}})]{KODAMA1985193}
Kodama, Y.
\newblock Normal forms for weakly dispersive wave equations.
\newblock {\em Physics Letters A} {\bf 1985}, {\em 112},~193--196.

\bibitem[Kodama(1985{\natexlab{b}})]{KODAMA1985245}
Kodama, Y.
\newblock On integrable systems with higher order corrections.
\newblock {\em Physics Letters A} {\bf 1985}, {\em 107},~245--249.

\bibitem[Camassa et~al.(1994)Camassa, Holm, and Hyman]{CAMASSA19941}
Camassa, R.; Holm, D.D.; Hyman, J.M.
\newblock A new integrable shallow water equation; Elsevier,  1994; Vol.~31,
  {\em Advances in Applied Mechanics}, pp. 1--33.

\bibitem[Charalampidis et~al.(2023)Charalampidis, Parker, Kevrekidis, and
  Lafortune]{Charalampidis_b_family_peakon_2023}
Charalampidis, E.; Parker, R.; Kevrekidis, P.; Lafortune, S.
\newblock The stability of the b-family of peakon equations.
\newblock {\em Nonlinearity} {\bf 2023}, {\em 36},~1192--1217.

\bibitem[Sulem and Sulem(1999)]{sulem_book}
Sulem, C.; Sulem, P.
\newblock {\em The {N}onlinear {S}chr\"odinger {E}quation}; Springer-Verlag
  (New York),  1999.

\bibitem[Chapman et~al.(2022)Chapman, Kavousanakis, Charalampidis, Kevrekidis,
  and Kevrekidis]{CHAPMAN2022133396}
Chapman, S.; Kavousanakis, M.; Charalampidis, E.; Kevrekidis, I.; Kevrekidis,
  P.
\newblock A spectral analysis of the nonlinear {S}chr{\"o}dinger equation in
  the co-exploding frame.
\newblock {\em Physica D: Nonlinear Phenomena} {\bf 2022}, {\em 439},~133396.

\bibitem[Chapman et~al.()Chapman, Kavousanakis, Charalampidis, Kevrekidis, and
  Kevrekidis]{kdvchapman}
Chapman, S.; Kavousanakis, M.; Charalampidis, E.; Kevrekidis, I.; Kevrekidis,
  P.
\newblock Self-similar blow-up solutions in the generalized {K}orteweg-de
  {V}ries equation: {S}pectral analysis, normal form and asymptotics.
\newblock  \href{http://xxx.lanl.gov/abs/2310.13770}{{\normalfont
  [arXiv:nlin.PS/2310.13770]}}.

\end{thebibliography}

\end{adjustwidth}
\end{document}